\documentclass[a4paper,11pt]{article}
\usepackage{jcappub} 

\usepackage{graphicx}
\usepackage{xcolor}
\usepackage[sort&compress]{natbib}
\usepackage{amsmath,amssymb,bm,bbm,slashed,subdepth}
\usepackage{enumerate}
\usepackage{epsfig, subfigure}
\usepackage{setspace}
\usepackage{booktabs, tabularx}
\usepackage{units}
\usepackage{placeins}
\usepackage{multirow}
\usepackage{mathtools}

\begin{document}

\begin{flushright}
		DESY-25-149\\
	\end{flushright}

\title{
Stellar Bounds on Light Spin-2 Particles in Bimetric Theories
}

\author[1]{Camilo Garc\'ia-Cely,}
\affiliation[1]{Instituto de F\'{i}sica Corpuscular (IFIC), Universitat de Val\`{e}ncia-CSIC, Parc Cient\'{i}fic UV, C/ Catedr\'{a}tico Jos\'{e} Beltr\'{a}n 2,
E-46980 Paterna, Spain}

\author[2]{Andreas Ringwald}
\affiliation[2]{Deutsches Elektronen-Synchrotron DESY, Notkestr. 85,
22607 Hamburg, Germany}

\abstract{
Using the bimetric formalism, we compute the production and emission rates of light spin-2 particles in non-degenerate stellar interiors through photoproduction and bremsstrahlung processes, including the effects of plasma screening. By comparing the resulting energy-loss rates with observational limits on stellar cooling, we derive bounds on the coupling strength and mass of the spin-2 particle. Assuming these particles are the dark matter of the Universe, the obtained constraints are competitive with existing astrophysical and cosmological limits, excluding a wide region of parameter space in the mass range 5--30~eV. 
}

\maketitle
\section{Introduction}

Multiple astrophysical and cosmological observations provide compelling evidence for the existence of dark matter.
However, its microscopic nature—including its spin—remains unknown.
Models in which dark matter consists of particles with spin $0$, $1/2$, or $1$ have been extensively explored since the 70s, as such fields can be easily embedded within extensions of the Standard Model (SM) of particle physics.
By contrast, formulating a consistent interacting theory for massive spin-2 fields poses significant theoretical challenges, primarily due to the emergence of unphysical ghost-like degrees of freedom.

This situation has changed over the past two decades with the development of bimetric gravity theories~\cite{Hassan:2011hr,Hassan:2011tf,Hassan:2012wr,deRham:2010kj}, which provide a ghost-free framework for describing a massive spin-2 particle coupled to gravity. In these theories, the gravitational sector is described by two interacting tensor fields: a massless one associated with the graviton of General Relativity, $\delta G_{\mu\nu}$, and a massive one, $\delta M_{\mu\nu}$, which couples universally to the SM through the stress-energy tensor. 
For a detailed review of these developments, see Ref.~\cite{Schmidt-May:2015vnx, deRham:2014zqa}.

These theoretical advances motivate the hypothesis that DM could be an elementary massive spin-2 particle arising from bimetric gravity~\cite{Maeda:2013bha,Aoki:2016zgp,Babichev:2016hir,Babichev:2016bxi}. In this framework, a single coupling between $\delta M$ and SM fields ensures that such particles interact only gravitationally at leading order, making them natural candidates for feebly interacting DM. Depending on the mass, several mechanisms have been proposed to account for the observed relic density of the spin-2 particle~\cite{Dubovsky:2004ud, Babichev:2016bxi, Marzola:2017lbt, Chu:2017msm, Cai:2021nmk, Kolb:2023dzp}. While cosmological aspects  have been investigated in previous studies, their implications in stellar environments have not been explored in comparable detail.

Closely following previous works~\cite{Cembranos:2017vgi, Garcia-Cely:2024ujr}, the purpose of this work is to examine the solar and stellar constraints on light massive spin-2 particles within the bimetric framework. Particles like $\delta M$ can be produced in stellar interiors through processes such as electron–nucleus or electron-electron bremsstrahlung as well as photoproduction. Once produced, they free-stream out of the star, carrying away energy and thereby modifying stellar evolution. These arguments have been successfully used~\cite{Raffelt:1996wa} to constrain other weakly interacting light particles such as axions, CP-even scalars, and dark photons. 
Here, we extend those analyses to the spin-2 case.

More precisely, by integrating the emission rates over realistic stellar models and incorporating plasma screening effects relevant in the dense cores of the Sun and horizontal-branch (HB) stars, we compute the corresponding energy-loss rates from spin-2 emission. 
These predictions are then compared with observational bounds on stellar cooling. 
The resulting constraints are competitive with, and in some cases stronger than, existing limits from cosmology, indirect dark matter searches, and laboratory tests of gravity. 
In particular, HB stars provide the most stringent bounds for spin-2 particle masses in the range of a few to several tens of electronvolts. 
We also compare the predicted solar spin-2 emission spectrum with the solar graviton spectrum recently calculated in Ref.~\cite{Garcia-Cely:2024ujr}, and discuss the prospects for detecting solar spin-2 fluxes in helioscope experiments such as CAST and IAXO through photon conversion in magnetic fields.

The paper is organized as follows. 
In Sec.~\ref{sec:bimetric}, we introduce the theoretical framework for massive spin-2 fields, their role as potential dark matter candidates and discuss parallels with axion phenomenology. 
Sec.~\ref{sec:emission} describes the production of light spin-2 particles in the solar plasma through bremsstrahlung and photoproduction and present the solar flux of spin-2 particles. 
Sec.~\ref{sec:bounds} presents the main results summarizing astrophysical and laboratory constraints on light spin-2 particles.
We present our  conclusions in Sec.~\ref{sec:conclusions}.
In this paper, we adopt Heaviside units, take $\hbar=c=1$, and utilize the Minkowski metric $\eta_{\mu\nu} = \text{diag}(+ - - -)$.

\section{Dark Matter in Bimetric Theories}
\label{sec:bimetric}

\subsection{Massive Spin-2 Fields}
\label{sec:spin2_Lagrangian}

Spin-2 particles are described by symmetric, traceless tensor fields $\delta M_{\mu\nu}$~\cite{Fierz:1939ix}. 
At the linear level, their dynamics and interactions with matter are captured by the Lagrangian
\begin{eqnarray}
\mathcal{L}[\delta M_{\mu\nu}]
&=&
\frac{1}{2}\partial_\rho \delta M_{\mu\nu}\partial^\rho \delta M^{\mu\nu}
-\frac{1}{2}\partial_\rho \delta M\,\partial^\rho \delta M
+\partial_\rho \delta M\,\partial_\nu \delta M^{\rho\nu}
-\partial_\rho \delta M_{\mu\nu}\partial^\nu \delta M^{\mu\rho}
\nonumber\\
&&
-\frac{1}{2}m^2\!\left(\delta M_{\mu\nu}\delta M^{\mu\nu}-(\delta M)^2\right)
+ (8\pi G')^{1/2}\,\delta M_{\mu\nu}\,T^{\mu\nu}\,,
\label{eq:Lspin2}
\end{eqnarray}
 The first line in Eq.~\eqref{eq:Lspin2} gives the canonical kinetic term of a massive spin-2 field, while the second line includes the Fierz–Pauli mass term~\cite{Fierz:1939ix} and its universal coupling to the SM energy–momentum tensor $T^{\mu\nu}$. 
 
For $m=0$ and $G'=G$, with $G$ the Newton constant, Eq.~\eqref{eq:Lspin2} coincides with the Lagrangian of linearized gravity describing the ordinary graviton, here denoted by $\delta G_{\mu\nu}$. 
In bimetric theories,   
 the total Lagrangian can be schematically written as
\begin{equation}
\mathcal{L}_{\rm bimetric} 
= \mathcal{L}[\delta M_{\mu\nu}]
+ \mathcal{L}[\delta G_{\mu\nu}]\Big|_{m=0, G'=G}
+ \mathcal{L}_{\rm int}[\delta G_{\mu\nu},\delta M_{\mu\nu}]\,,
\label{eq:Lbimetric}
\end{equation}
where the second term arises from the Einstein–Hilbert action and describes massless gravitons, while the third term, $\mathcal{L}_{\rm int}$, encodes nonlinear interactions between the two tensor fields, which scale with negative powers of the ratio $G'/G$~\cite{Babichev:2016bxi}. 
As will be shown below, the parameter region relevant for stellar production of spin-2 particles corresponds to $G' \gg G$, in which case these nonlinear terms are strongly suppressed. 
Although they are essential for the theoretical consistency of bimetric gravity—being constructed to eliminate the Boulware–Deser ghost~\cite{Boulware:1972yco} and to guarantee the stable propagation of both spin-2 modes—they have no practical impact on the emission or decay processes considered below.

We will be interested only in processes involving a single spin-2 particle, either in its production or decay. For such processes, the corresponding transition amplitude can be cast as
\begin{equation}
{\cal M}(\lambda)={\cal M}_{\mu\nu}\,\epsilon^{\mu\nu}_{(\lambda)}(p),
\label{eq:AmplitudeMassive}
\end{equation}
where $p=(\omega,\mathbf{p})$ and $\lambda$ denote respectively the momentum and polarization state for the spin-2 particle, while  $\epsilon^{\mu\nu}_{(\lambda)}$ is the corresponding spin-2 polarization tensor. They are given by~\cite{Hagiwara:2008jb, Christensen:2013aua}
\begin{align}
\epsilon^{\mu\nu}_{\pm2}(p)=\epsilon^\mu_{\pm}(p)\,\epsilon^\nu_{\pm}(p),
&&\text{with}&&
\epsilon^\mu_{\pm}(p)=\frac{1}{\sqrt{2}}
\begin{pmatrix}
0\\
\mp\cos\theta\cos\phi+i\sin\phi\\
\mp\cos\theta\sin\phi-i\cos\phi\\
\pm\sin\theta
\end{pmatrix},
\label{eq:PolarizationsMassless}
\end{align}
and three additional polarizations of a spin-2 multiplet, which can be constructed as symmetric products of the polarization vectors of a massive spin-1 field. Concretely,   
\begin{align}
\epsilon^{\mu\nu}_{\pm1}(p) &= \frac{1}{\sqrt{2}}\left[
\epsilon^\mu_0(p)\,\epsilon^\nu_{\pm}(p)
+ \epsilon^\mu_{\pm}(p)\,\epsilon^\nu_0(p)
\right],
\label{eq:extraspins_1}
\ \ \text{and}
\\
\epsilon^{\mu\nu}_{0}(p) &= \frac{1}{\sqrt{6}}\left[
\epsilon^\mu_-(p)\,\epsilon^\nu_+(p)
+ \epsilon^\mu_+(p)\,\epsilon^\nu_-(p)
+ 2\,\epsilon^\mu_0(p)\,\epsilon^\nu_0(p)
\right]\,,
\label{eq:extraspins_0}
\end{align}
where 
\begin{align}
\epsilon^\mu_0(p) = \frac{1}{m}
\begin{pmatrix}
|\mathbf{p}| \\[4pt]
\omega \sin\theta \cos\phi \\[4pt]
\omega \sin\theta \sin\phi \\[4pt]
\omega \cos\theta
\end{pmatrix}\,,
&&
\text{with}
&&
p^\mu=
\begin{pmatrix}
\omega\\
|\mathbf{p}|\sin\theta\cos\phi\\
|\mathbf{p}|\sin\theta\sin\phi\\
|\mathbf{p}|\cos\theta
\end{pmatrix}.
\end{align}
These polarization tensors form an orthonormal basis,  and satisfy the transversality and tracelessness conditions
\begin{align}
\epsilon^{\mu\nu}_{(\lambda)}(p)\,\epsilon^{*}_{(\lambda')\,\mu\nu}(p) = \delta_{\lambda \lambda'}\,,
&&
p_\mu\, \epsilon^{\mu\nu}_{(\lambda)}(p) = 0\,,
&&
\eta_{\mu\nu}\,\epsilon^{\mu\nu}_{(\lambda)}(p) = 0\,.
\label{eq:constraints}
\end{align}
We focus on spin-2 particles much lighter than any other relevant energy scale and therefore work in the limit $m \to 0$.\footnote{This limit has been shown to be technically natural, see e.g.~\cite{deRham:2012ew}.} As will be discussed below, this is not equivalent to setting $m = 0$.
Then, to compute the transition amplitudes for the processes of interest in this work, we implement Eq.~\eqref{eq:Lbimetric} in \texttt{FeynRules}~\cite{Alloul:2013bka}, which generates the model files required by \texttt{CalcHEP}~\cite{Belyaev:2012qa}.  
The resulting symbolic output is then processed with \texttt{Package-X}~\cite{Patel:2016fam} to obtain explicit expressions for the matrix element ${\cal M}^{\mu\nu}$. We then check that these expressions satisfy the transversality condition
\begin{equation}
p^\mu {\cal M}_{\mu\nu}=0,
\label{eq:Conservation}
\end{equation}
which follows from energy-momentum conservation for $m\to0$.

We also note  that for processes involving a single external $\delta M$ field, the Feynman rules derived from Eq.~\eqref{eq:Lbimetric} coincide with those of the ordinary graviton (see, e.g., Refs.~\cite{Choi:1994ax,Donoghue:1994dn}), up to an overall rescaling of the coupling constant. 
We may therefore also employ the interaction rules summarized in the Supplemental Material of Ref.~\cite{Garcia-Cely:2024ujr}.

 \subsection{Light Spin-2 Particles as Dark Matter}
\label{sec:DM}

Since the massive spin-2 particle carries no SM quantum numbers, it decays  into all kinematically allowed SM channels. This universal coupling pattern is shared with other spin-2 frameworks such as Kaluza–Klein dark matter (see e.g. Ref.~\cite{Han:1998sg}).   A key distinction of bimetric theories, however, is that the massive eigenstate $\delta M$ cannot decay into the massless mode. In other words, the decay $\delta M \to \delta G\,\delta G$ is forbidden by the structure of the bimetric potential~\cite{Babichev:2016hir}, which does not generate interaction vertices involving one massive and two massless spin-2 fields. As a result, no gravitational-wave or graviton production accompanies the decay of our massive spin-2 field~(see e.g. Refs.~\cite{Strumia:2025dfn, Dunsky:2025pvd, Cembranos:2025uhe} for recent studies exploring this channel).
Using Eq.~\eqref{eq:AmplitudeMassive} as explained above, for keV-scale masses or below we find the decay widths~\cite{Babichev:2016bxi}
\begin{equation}
\Gamma
=
\frac{m^{3} G'}{640\pi^2 } \times
\begin{cases}
\frac{1}{2}, & \text{for }\delta M\to\gamma\gamma,\\[4pt]
\frac{3}{16}, & \text{for }\delta M\to\nu\bar{\nu}.\\[4pt]
\end{cases}
\label{eq:GammaDeltaM}
\end{equation}
As will be shown below, these decay channels make the scenario experimentally testable, providing complementary probes to the stellar emission processes analyzed in this work. 

Concerning DM  production, several mechanisms have been proposed to explain the relic abundance of a massive spin-2 particle, including gravitational production during reheating, freeze-in via the gravitational portal, and misalignment-like scenarios~\cite{Dubovsky:2004ud, Marzola:2017lbt, Kolb:2023dzp, Cai:2021nmk, Babichev:2016bxi, Chu:2017msm}. 
In the present work, however, we remain agnostic regarding the specific origin of the relic density. 
Our focus is instead on the phenomenology of stellar environments, where the production and emission of spin-2 particles can lead to observable consequences independent of the particular cosmological history.

\subsection{Parallels with Axion Phenomenology}
\label{sec:photoconversion}

Axions are pseudoscalar particles originally introduced to solve the strong CP problem~\cite{Peccei:1977hh,  Weinberg:1977ma,Wilczek:1977pj, Vafa:1984xg} and are among the most studied candidates for light dark matter~\cite{Preskill:1982cy,Abbott:1982af, Dine:1982ah}.  
They couple to two photons through the interaction  
\begin{equation}
\mathcal{L}_{a\gamma\gamma} = -\frac{1}{4} g_{a\gamma\gamma} a F_{\mu\nu} \tilde{F}^{\mu\nu}
= g_{a\gamma\gamma}\, a\, \mathbf{E}\!\cdot\!\mathbf{B}\,,
\label{eq:Laxion}
\end{equation}
where $g_{a\gamma\gamma}$ is the axion–photon coupling constant, $a$ is the axion field, and $\mathbf{E}$ and $\mathbf{B}$ denote the electric and magnetic fields.  
This interaction allows axions to decay into two photons in analogy with the decay of spin-2 particles discussed above, with a width~\cite{Arias:2012az}
\begin{equation}
\Gamma ({a\to\gamma\gamma})
= \frac{g_{a\gamma\gamma}^2 m^3}{64\pi}.
\label{eq:Gamma_a}
\end{equation}

\medskip
\noindent
The same coupling in Eq.~\eqref{eq:Laxion} also enables the production of axions in stars through the Primakoff process~\cite{Primakoff:1951iae,Dicus:1978fp}, in which thermal photons convert into axions in the Coulomb fields of charged particles in stellar plasmas. As we will see below, spin-2 particles are produced in stars through analogous mechanisms, albeit with important differences. 

The axions produced in the Sun are the target of helioscope experiments~\cite{Sikivie:1983ip}, which aim to detect them through their conversion into photons via the inverse Primakoff effect.  
In a strong transverse magnetic field, the interaction term in Eq.~\eqref{eq:Laxion} induces mixing between axion and photon states, leading to a conversion probability 
\begin{equation}
P_{a\to\gamma} = \frac{1}{4} g_{a\gamma\gamma}^{2} B^{2} L^{2}
\,\mathrm{sinc}^2\!\left(\frac{qL}{2}\right),
\label{eq:PrimakoffConversion}
\end{equation}
where $B$ is the magnetic-field strength, $L$ the length of the magnet, and 
$q = m_a^2 / (2\omega)$ is the momentum transfer between the photon and the axion. In this way, the CERN Axion Solar Telescope (CAST)~\cite{CAST:2017uph} has established the strongest laboratory limits on the axion–photon coupling $g_{a\gamma\gamma}$.

Likewise, a light spin-2 particle can convert into a photon in the presence of an external magnetic field due to the coupling between the electromagnetic field strength tensor and the spin-2 field in Eq.~\eqref{eq:Lspin2}. This effect, analyzed in detail in Ref.~\cite{Biggio:2006im}, originates from the interaction term
\begin{equation}
\mathcal{L}_{\delta M\gamma\gamma} = (8\pi G')^{1/2}\, \delta M_{\mu\nu} F^{\mu\alpha} F^{\nu}{}_{\alpha} 
\label{eq:Lmix}
\end{equation}
When an electromagnetic wave propagates through a uniform magnetic field, one of the field tensors in Eq.~\eqref{eq:Lmix} is replaced by the external field, inducing off-diagonal terms in the equations of motion that mix the photon and spin-2 states.  
The resulting system behaves as a two-level oscillation, analogous to axion–photon conversion.  
The corresponding conversion probability is 
\begin{equation}
P_{\delta M_\lambda\to\gamma} = 8\pi G'L^2 |\vec{\epsilon}^{\,\,*} \cdot \epsilon_{(\lambda)} \cdot(\vec{B}\times \hat{p}) |^2 
\,\mathrm{sinc}^2\!\left(\frac{qL}{2}\right),
\label{eq:conversionM}
\end{equation}
where $\hat{p}$ is the direction of propagation, $\vec{\epsilon} $ denotes the polarization vector of the photon and $\epsilon_{(\lambda)}$ is the matrix describing the polarization of the spin-2 particle. Here we assume that the photon and the spin-2 particle propagate in the same direction. See Ref.~\cite{Domcke:2025qlw} for a recent discussion on relaxing this assumption.

According to Eq.~\eqref{eq:conversionM}, the modes $\lambda = \pm 2$ produce circularly polarized photons with helicity $\pm 1$, independently of the orientation of the magnetic field, whereas the scalar mode $\lambda = 0$ produces linearly polarized photons whose polarization vector lies perpendicular to the plane defined by the magnetic field and the direction of propagation,  see Table~\ref{table:br}
For comparison, axions produce linearly polarized photons with polarization parallel to this plane~\cite{Raffelt:1987im}.
Singling out the spin-2 polarization but summing over that of the photon, we find\footnote{
The probability $P_{\delta M_{\lambda=0}\rightarrow \gamma}$ in Eq.\eqref{eq:conversion0} differs from the corresponding expression in Ref.~\cite{Biggio:2006im}.
We trace this discrepancy to their normalization of the spin-0 mode in Eq.~(A7), which is larger by a factor of two than what is implied by their Eq.~(A9).
This propagates to their light-shining-through-a-wall probability, giving a factor $49/9$ in Eq.~(69) instead of $16/9$, which is the factor we obtain following the procedure outlined here.
}
\begin{align}
P_{\delta M_{\lambda=\pm 2}\rightarrow \gamma}  = 3 P_{\delta M_{\lambda= 0}\rightarrow \gamma}  = 4\pi G' B^2 L^2 \, \text{sinc}^2\!\left(\frac{qL}{2}\right), 
&&\text{and}&&
P_{\delta M_{\lambda=\pm 1}\rightarrow \gamma}& = 0\,.
\label{eq:conversion0}
\end{align}
\begin{table}[t]
\centering
\begin{tabular}{c|cc|cc}
$P_{\delta M\rightarrow \gamma}/4\pi G' B^2 L^2 \, \text{sinc}^2(qL/2)$ & $\gamma_\parallel$&   $\gamma_\perp$ & $\gamma_{+}$& $\gamma_{-}$\\\hline
$\lambda=2$ & $\frac{1}{2}$ & $\frac{1}{2}$  &1 & 0 \\ 
$\lambda=1$ &  $0$  & $0$ &  $0$  & $0$ \\ 
$\lambda=0$ &$0$ & $\frac{1}{3}$  & $\frac{1}{6}$ & $\frac{1}{6}$\\
$\lambda=-1$ & $0$ & $0$  &$0$ &$0$\\ 
$\lambda=-2$ & $\frac{1}{2}$ & $\frac{1}{2}$  &  0 & 1 \\\hline 
\end{tabular}
\label{table:br}
\caption{Branching ratios associated with the conversion probabilities depending on the polarization of the produced photon.}
\end{table}
Interestingly, the conversion probability for $\lambda=\pm2$ and $G'=G$ is the same as that of the inverse Gertsenshtein effect~\cite{Gertsenshtein,Boccaletti1970,Raffelt:1987im,Ejlli:2019bqj,Domcke:2020yzq}, by which gravitational waves convert into electromagnetic waves in an external magnetic field.

\subsection{The van Dam–Veltman–Zakharov (vDVZ) Discontinuity and the  Vainshtein Mechanism.} 

For two non-relativistic particles of masses $m_i$ and $m_j$, the exchange of the massless and  massive spin-2 fields in Eq.~\eqref{eq:Lbimetric} induces the potential~\cite{Aoki:2016zgp}
\begin{equation}
V(r) = - \frac{m_i m_j}{r}\Big[\, G + \frac{4}{3}\, G' e^{- m r} \Big]\,.
\label{eq:yukawa}
\end{equation}
The factor $4/3$ is the well-known vDVZ discontinuity~\cite{vanDam:1970vg,Zakharov:1970cc}, which originates from the fact that, in the limit $m \to 0$, the potential associated with the exchange of the spin-0 component~\eqref{eq:extraspins_0}  does not vanish. This contrasts with the vector polarization modes, which according to Eq.~\eqref{eq:extraspins_1} behave as $\epsilon^{\mu\nu}_{\pm 1} \propto (p^\mu\,\epsilon^\nu_{\pm}(p) + \epsilon^\mu_{\pm}(p)\,p^\nu)$ in the limit $m \to 0$, and necessarily decouple as a consequence of the conservation condition in Eq.~\eqref{eq:Conservation}.
This discontinuity will reappear in various forms throughout the results presented below.

The vDVZ discontinuity shows that a theory containing only a massive spin-2 field, i.e. without the accompanying massless graviton, fails to reproduce the Newtonian potential in the limit $m \to 0$. After this was identified as a failure of massive gravity, Vainshtein pointed out that this effect disappears if  nonlinear effects are important~\cite{Vainshtein:1972sx}. Concretely, he showed that the additional scalar mode responsible for the discontinuity becomes strongly coupled near massive sources, below a characteristic distance known as the Vainshtein radius, $r_V$. Inside this radius, nonlinear interactions dominate and screen the extra polarization, effectively restoring the Newtonian potential when $m\to0$.

In bimetric theories, the Vainshtein screening mechanism generalizes naturally, defining the regime where nonlinear effects in the massive mode $\delta M_{\mu\nu}$ become significant. The corresponding Vainshtein radius is given by~\cite{Babichev:2016bxi}
\begin{equation}
r_V \approx \left[\frac{2(G'+G)M}{m^2}\right]^{1/3}\,.
\label{eq:rV}
\end{equation}
Throughout this work, we will focus on physical situations occurring well outside $r_{V}$, where the dynamics are accurately described by the linearized theory and the potential in Eq.~\eqref{eq:yukawa} remains valid. In particular, the characteristic $4/3$ enhancement correctly encodes the contribution of the residual scalar polarization. The corresponding parameter space relevant for comparing stellar bounds with other constraints is detailed in Sec.~\ref{sec:fifthforce}.

\section{Stellar Flux of Light Spin-2  Particles }
\label{sec:emission}
\subsection{Solar Plasma Model}

We will model the Sun as a nonrelativistic, fully ionized plasma composed primarily of electrons and nucleons following Maxwell–Boltzmann distributions. The temperature and density profiles are taken from the B16-GS98 solar model~\cite{Vinyoles:2016djt}, which provides accurate  quantities relevant to the emission processes. Elements heavier than Helium are neglected as their contribution to the total emission is insignificant. The plasma is treated as nondegenerate, a condition verified by comparing the electron Fermi energy with the thermal energy across the solar interior. Under these conditions, the relevant scattering processes —electron–ion, electron–electron, and photon–charged particle interactions— can be evaluated using the Born approximation within the inner 85\% of the Sun~\cite{Garcia-Cely:2024ujr}.

Screening effects are incorporated through a Yukawa potential between charged particles characterized by the Debye–Hückel scale~\cite{1954AuJPh...7..373S,Raffelt:1985nk}
\begin{equation}
\kappa = \left[\frac{4\pi\alpha(n_e + \sum_Z Z n_Z)}{T}\right]^{1/2},
\label{eq:kappa}
\end{equation}
which accounts for collective Coulomb interactions in the plasma. This screening length regulates the long-range divergence of the Coulomb potential and sets an effective infrared cutoff in the calculation of bremsstrahlung and photoproduction rates to be studied below. 
Furthermore, since the solar plasma is nonrelativistic and $T \gg \omega_{\mathrm{pl}}$, where $\omega_{\mathrm{pl}}$ is the plasma frequency, the photon dispersion relation introduces only minor corrections. This treatment ensures that all microscopic emission processes are finite and well-defined throughout the solar volume, allowing for a consistent computation of the total emission rate.

Under these conditions, the differential emission rate per unit volume for the production of a single spin-2 particle in the collision of two particles is given by  
\begin{equation}
\!\!\!\frac{\mathrm{d}\Gamma}{\mathrm{d}\omega\, \mathrm{d}V}\!\left(1+2\rightarrow \delta M_\lambda+\cdots\right)
= \int dn_1\, dn_2\, |{\cal M}(\lambda)|^2\, \mathrm{d(PS)}\,
(2\pi)^4\,\delta^{(4)}\!(p_1+p_2-p-\sum_k p_k),
\label{eq:GammaMassive}
\end{equation}
where the phase-space element, isolating the contribution of the emitted spin-2 particle, reads  
\begin{equation}
\mathrm{d(PS)} = \frac{\omega^2\, \mathrm{d}\Omega_{\mathbf{p}}}{(2\pi)^3\,2\omega}
\prod_k \frac{\mathrm{d}^3p_k}{(2\pi)^3\,2E_k}.
\label{eq:PhaseSpaceMassive}
\end{equation}

\begin{figure}[t]
\includegraphics[height=0.36\textheight]{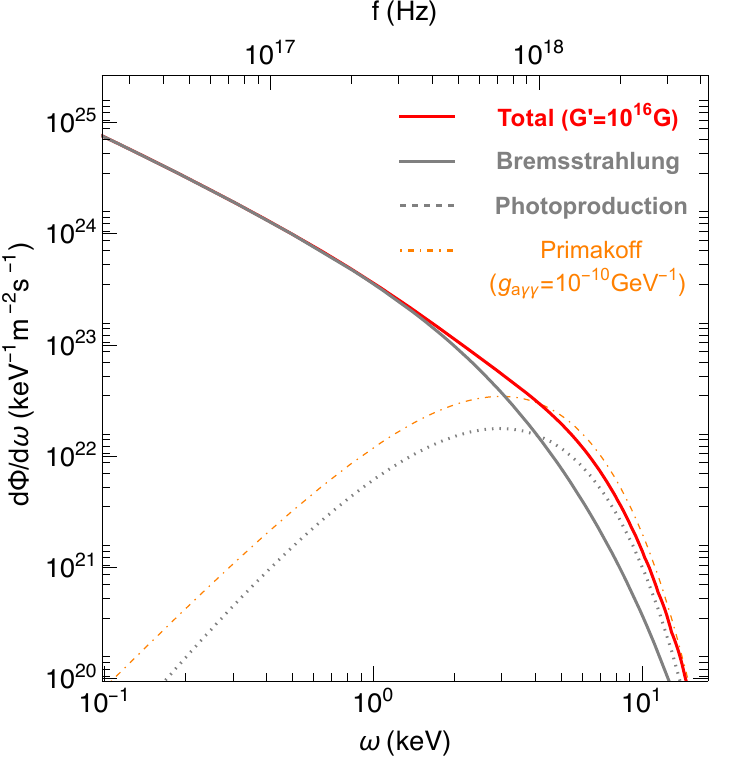 }
\includegraphics[height=0.36\textheight]{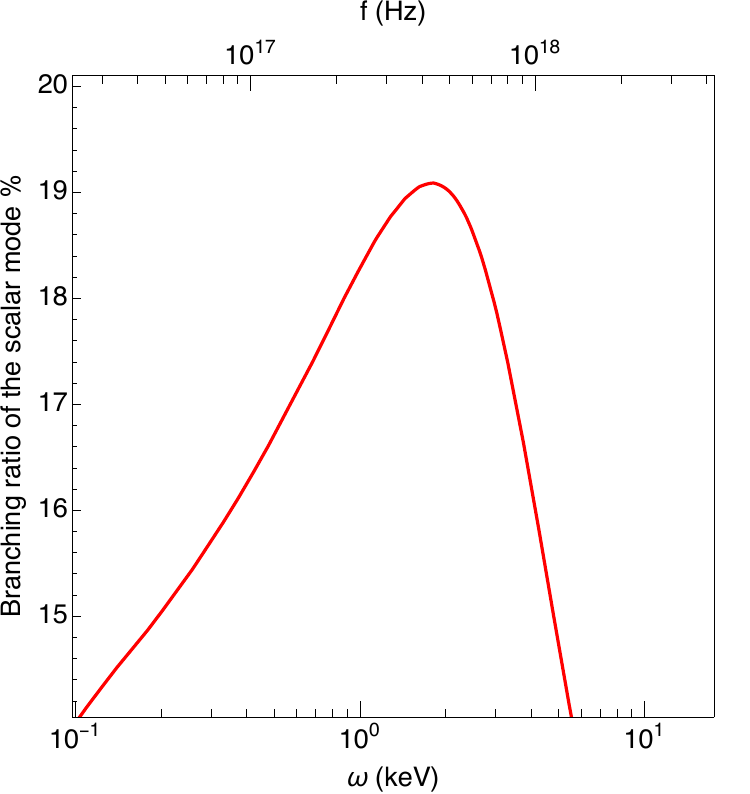}

\caption{ \emph{Left panel:} 
Differential flux at Earth of the number of spin-2 particles produced in the Sun via bremsstrahlung and photoproduction, for $G'/G=10^{16}$, compared to the corresponding flux of axions produced in the Sun via the Primakoff process, for $g_{a\gamma\gamma}=10^{-10}\,\text{GeV}$.
\emph{Right panel:} 
Branching ratio of the scalar polarization mode ($\lambda = 0$) relative to the total solar emission of spin-2 particles. 
The result shows that the emission is dominated by the tensor components ($\lambda = \pm 2$), and consequently, the spectrum of light spin-2 particles produced in the solar plasma closely follows that of ordinary gravitons, differing only by an overall normalization factor $G'/G$.
}
\label{fig:spectrum}
\end{figure}

The differential flux of spin-2 particles arriving at Earth, defined per unit area, per unit time, and per unit energy, is then obtained by thermally averaging the emission rates throughout the solar interior
\begin{equation}
\frac{\mathrm{d}\Phi}{\mathrm{d}\omega}
= \frac{1}{4\pi d_\odot^2}
\int_{\mathrm{Sun}} \!\!\mathrm{d}^3\mathbf{r}\,
\sum_i 
\left\langle 
\frac{\mathrm{d}\Gamma^{(i)}(\mathbf{r})}{\mathrm{d}\omega\,\mathrm{d}V}
\right\rangle.
\label{eq:collisions}
\end{equation}
We assume spherical symmetry for the solar integration, and the index $i$ runs over the two relevant production mechanisms: photoproduction and bremsstrahlung.  
Before discussing these processes in detail, we note that the resulting flux, $\mathrm{d}\Phi/\mathrm{d}\omega$, for $G' = 10^{16} G$, is displayed in the left panel of Fig.~\ref{fig:spectrum}.

\begin{table*}[t]
\!\!\!\!
  \begin{tabular}{|c|c|c|}
    \hline
\multirow{2}{*}{\bf{Collision}} &
\multirow{2}{*}{$\mathbf{\lambda }$} & 
\multirow{2}{*}{$\frac{\mathrm{d}\sigma v}{\mathrm{d}\omega}$}
\\ && \\ \hline
{\small Photo-production}
       &
       $\pm 2$
      &
      $  G' Z^2  \alpha \pi\,  \delta(\omega-p_i) \int \mathrm{d}\!\cos\theta\cot^2\!\frac{\theta}{2} [1+\cos^2\theta] F(\theta) $
\\
\multirow{2}{*}{ $\gamma Z\to  Z \,\delta M_\lambda$}
&$\pm1$&0
\\
& $0$ & 
$  \frac{4}{3}    G'  Z^2 \alpha \pi  \,   \delta(\omega-p_i)\int \mathrm{d}\!\cos\theta\cot^2\!\frac{\theta}{2}\sin^4\!\frac{\theta}{2}F(\theta) $
\\\hline
{\small Bremsstrahlung}
     &
      $\pm 2$ 
       &
     $
\frac{32   G'  Z^2 \alpha^2 p_i }{15\omega} \left(\frac{1}{m_e}+\frac{1}{m_Z} \right) \left(3 (1+\xi ^2)
{L}
+10 \xi +{\cal O}(\xi_s^2)
\right)
  $ 
\\
\multirow{2}{*}{ $eZ\to eZ \,\delta M_\lambda $}&$\pm 1$&0
\\
& $0$ 
&
$\frac{16   G'  Z^2\alpha^2 p_i }{45\omega}  \left(\frac{1}{m_e}+\frac{1}{m_Z} \right) \left((1+\xi ^2)
{L}
+30 \xi  +{\cal O}(\xi_s^2) 
\right)
$
\\\hline
{\small Bremsstrahlung}
     &
   $\pm 2$ 
       &
      $ \frac{
    32  G' \alpha^2 p_i }{15\omega m_e}\left(
     \left(6 (1+\xi ^2)-\frac{3 (1-\xi ^2)^4+7 (1-\xi^4)^2}{2
   (1+\xi ^2)^3}\right)
{L}+
20 \xi -\frac{6 \xi  (1+\xi ^4)}{(1+\xi ^2)^2}
 +{\cal O}(\xi_s^2)
 \right) 
  $ 
\\
\multirow{2}{*}{$e e\rightarrow e e\ \delta M_\lambda$}&$\pm 1$&0
\\ 
& $0$ 
&
  $
     \frac{32 G'  \alpha^2  p_i}{15\omega m_e}\left(
\left(\frac{1}{3} (1+\xi ^2)-\frac{(1-\xi ^2)^4+29 (1-\xi ^4)^2}{12
   (1+\xi ^2)^3}\right)
{L}
 +\frac{29 \xi }{3}+\frac{2 \xi ^3}{3 (1+\xi ^2)^2}
   +{\cal O}(\xi_s^2)
 \right)
  $ 
\\\hline
   \end{tabular}
    \caption{Cross section for the emission of a spin-2 particle of negligible mass, energy $\omega$, and polarization $\lambda$, coupled to ordinary particles via Eq.~\eqref{eq:Lbimetric}. The incoming particles have momentum $p_i$ and total kinetic energy $E_i$. The expressions for $\lambda=\pm 2$ coincide with the corresponding emission rates of ordinary gravitons if $G' = G$~\cite{Garcia-Cely:2024ujr}. For photoproduction from electrons, the reported expression applies with $Z=1$.
    }
    \label{table:processes}  
\end{table*}

\subsection{Contribution from Photoproduction}

This is the production of a spin-2 particle from photon scattering off a charged particle in the plasma, more precisely, $\gamma Z \to Z\,\delta M$ or $\gamma e^- \to e^-\,\delta M$. This process is analogous to the Primakoff and Compton mechanisms for axion production in stars. 
For axions, these correspond to distinct interactions—the Primakoff process involving the axion–photon coupling and the Compton process involving the axion–electron coupling. 
In contrast, for spin-2 particles both contributions arise from the same universal coupling to the energy–momentum tensor, see Eq.~\eqref{eq:Lspin2}, and thus cannot be treated as separate processes. Using the method described above,  Eq.~\eqref{eq:GammaMassive} gives
\begin{equation}
\frac{{\rm d}\Gamma}{{\rm d}\omega\,{\rm d}V}
\Bigg|_{\mathrm{Photoproduction}}
=  n_{\gamma}\, n_e\,  \frac{{\rm d}\sigma}{{\rm d}\omega}\Bigg|_{\gamma e \rightarrow e\,\delta M_{\lambda}}+\sum_Z n_{\gamma}\, n_Z\,  \frac{{\rm d}\sigma}{{\rm d}\omega}\Bigg|_{\gamma Z \rightarrow Z\,\delta M_{\lambda}},
\label{eq:rateMassive} 
\end{equation}
where ${\rm d}\sigma/{\rm d}\omega$ denotes the differential cross section for each helicity mode, reported in Table~\ref{table:processes} (with $v=c$ due to the initial-state photon).  Here $n_\gamma$, $n_e$ and $n_Z$ are the photon, electron and relevant nucleon density.  As discussed above,  
the mode with $\lambda=\pm1$ decouples, leaving only $\lambda=\{0,\pm2\}$ as dynamically relevant.

The total rate exhibits a divergence at $\theta=0$. The divergence is regularized  by including plasma effects~\cite{Raffelt:1985nk, Hoof:2021mld}. After adding over all channels, this results in the form factor 
\begin{align}
F(\theta) &= \frac{|\mathbf{q}|^2}{\kappa^2 + |\mathbf{q}|^2} = \frac{\left(2 \omega  \sin \!\frac{\theta }{2}\right)^2}{\kappa ^2+\left(2 \omega  \sin \!\frac{\theta }{2}\right)^2}, 
\end{align}
where $\kappa$ is the screening scale in Eq.~\eqref{eq:kappa}, $\mathbf{q}$ is the momentum transfer in the collision and $\theta$ is the scattering angle in the center of mass frame.

\subsection{Contribution from Bremsstrahlung }

This process corresponds to the emission of a spin-2 particle in $eZ$ and $ee$ collisions, $eZ \to eZ\,\delta M_\lambda$ and $ee \to ee\,\delta M_\lambda$.  
Using the method described above, Eq.~\eqref{eq:GammaMassive} gives
\begin{equation}
\frac{{\mathrm d}\Gamma}{{\mathrm d}\omega\,{\mathrm d}V}\Bigg|_{\mathrm{Bremsstrahlung}}
=\frac{1}{2}n_e^2\,\frac{{\mathrm d}\sigma v}{{\mathrm d}\omega}\Bigg|_{ee\to ee\,\delta M}
+\sum_Z n_e n_Z\,\frac{{\mathrm d}\sigma v}{{\mathrm d}\omega}\Bigg|_{eZ\to eZ\,\delta M},
\label{eq:BremsRateMassive}
\end{equation}
where the non-relativistic cross sections are reported in Table~\ref{table:processes}, without assuming $m_Z \gg m_e$. See Appendix \ref{sec:appA} for details.
The kinematics are conveniently described by  
\begin{align}
\xi = \frac{p_f}{p_i}, \qquad 
\omega = E_i (1-\xi^2)\,,
\end{align}
where the mass of the emitted spin-2 particle is neglected. The emission of spin-2 particles of very low-energy corresponds to $\omega \to 0$, or equivalently $\xi \to 1$. 
In this limit, the differential cross sections develop a logarithmic divergence, $\log\omega$.  
Plasma effects, however, regularize this behavior through an infrared cutoff associated with the screening scale defined in Eq.~\eqref{eq:kappa}.   To implement this, we closely follow Ref.~\cite{Raffelt:1985nk}, where screening is incorporated by introducing an effective photon mass equal to the Debye–Hückel scale $\kappa$,\footnote{A more systematic treatment can be formulated within the framework of thermal field theory, where the inclusion of hard thermal loop self-energies in the propagators automatically regularizes the infrared divergences over the entire frequency spectrum, as demonstrated for axions in Ref.~\cite{Altherr:1992mf}. 
In this approach, the screening parameter $\kappa^2$ naturally arises as the photon self-energy evaluated in the static limit ($\omega \to 0$)~\cite{Kapusta:1989tk,Raffelt:1996wa}. } appropriate for the slow, non-relativistic plasma components of stellar interiors.  
In practice, this results in a regulated logarithm of the form  
\begin{align}
L = \log \sqrt{\frac{(1+\xi)^2+\xi _s^2}{(1-\xi )^2+\xi _s^2}}, \qquad \text{where} \qquad \xi_s = \frac{\kappa}{p_i}\,.
\end{align}
This logarithm explicitly appears in the differential non-relativistic cross sections  reported in Table~\ref{table:processes}, which   
have been expanded in $\xi_s$. The complete expressions are given in Appendix~\ref{sec:appA}.  
Once again, the $\lambda=\pm1$ modes decouple, leaving only $\lambda=\{0,\pm2\}$ as physically relevant.

Before discussing the results, let us  note that our calculations assume the validity of the Born approximation, which remains accurate for $ee$ and $eZ$ scattering processes throughout most of the solar interior, except in the outermost layers, roughly beyond $0.85\,R_\odot$~\cite{Hoof:2021mld,Garcia-Cely:2024ujr}.  
Because bremsstrahlung emission is dominated by the dense central regions of the Sun, neglecting non-perturbative corrections to the spin-2 production rate is well justified.  

Fig.~\ref{fig:spectrum} shows the differential flux of spin-2 particles emitted by the Sun as a function of energy, for a benchmark coupling $G' = 10^{16}G$.  
The total flux (solid red line) is obtained by summing the individual contributions from photoproduction and bremsstrahlung processes.
The former, shown as the dotted gray curve, accounts for the high-energy part of the spectrum, whereas bremsstrahlung, shown as a solid gray line, dominates at lower photon energies, due to the regulated infrared divergence.   
For comparison, the expected solar axion flux from the Primakoff process is also shown (dash-dotted orange line) for a reference coupling $g_{a\gamma\gamma} = 10^{-10}\,\mathrm{GeV^{-1}}$.

\subsection{Comparison with the Solar Gravitational-Wave Spectrum}
\label{sec:comparison}

The right panel of Fig.~\ref{fig:spectrum} displays the branching ratio of the scalar polarization mode ($\lambda=0$) relative to the total emission of solar spin-2 particles. Across the energy range of interest, this fraction never exceeds about 20\%. 
With the emission dominated by $\lambda=\pm 2$ modes, the spectrum of light spin-2 particles produced in the solar plasma follows closely the same spectral shape as that of ordinary gravitons emitted by the Sun, differing only by an overall normalization factor proportional to $G'/G$. The only caveat is that this correspondence holds provided the mass of the spin-2 particle is sufficiently small compared with the solar temperature, ensuring kinematics remain effectively identical to the massless case.

In fact, the expressions reported in Table~\ref{table:processes} for $\lambda=\pm2$ exactly match those found for gravitons in Ref.~\cite{Garcia-Cely:2024ujr}. This work recently revisited the full solar GW spectrum, extending Weinberg’s classical result on the spectrum of soft gravitons~\cite{Weinberg:1965nx} to the full solar frequency range. That work provided a comprehensive treatment of both microscopic and macroscopic sources of GW emission in the solar plasma. The microscopic component, originating from particle collisions such as bremsstrahlung and photoproduction, was computed using realistic temperature and density profiles and the same screened plasma description adopted here. 
A number of earlier works have investigated related emission processes.  
Weinberg’s original calculation~\cite{Weinberg:1965nx,Weinberg:1972kfs} established that soft graviton  bremsstrahlung from particle collisions yields a nearly flat power spectrum $\omega\mathrm{d}\Phi/\mathrm{d}\omega$, consistent with the classical quadrupole emission formula in the appropriate limit~\cite{Gould:1985,Steane:2023gme}.  
Subsequent studies refined this result by extending the analysis beyond the soft limit, applying it to $eZ$ and $ee$ collisions~\cite{Barker:1969jk,Papini:1977fm,Gould:1981an,Gould:1985}, though typically neglecting plasma screening or assuming infinitely heavy scatterers.  
Our present calculation, for $\lambda=\pm 2$,  reproduces these results in the corresponding limits while incorporating a consistent treatment of screening effects.
Likewise, for photoproduction, our rates  for $\lambda=\pm 2$ reduce to those reported in Ref.~\cite{Voronov:1973kga} when screening is neglected, as expected.

\section{Astrophysical and Laboratory Constraints on Light Spin-2 Particles}
\label{sec:bounds}

\begin{figure}[t]
\includegraphics[width=\textwidth]{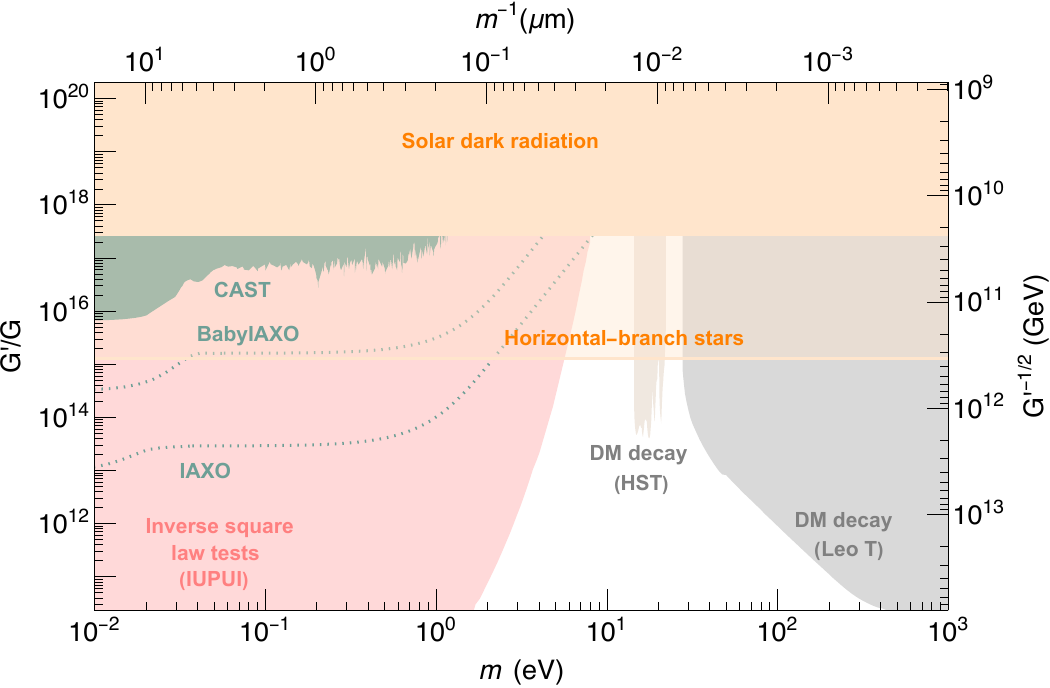}
\caption{
Constraints on the coupling ratio $G'/G$ as a function of the mass, $m$, of the spin-2 particle. 
The stellar energy-loss bounds derived in this work from the Sun and HB stars are shown in orange, together with limits on dark matter decay into photons from the Leo~T dwarf galaxy~\cite{Wadekar:2021qae} and from ultraviolet observations of dwarf galaxies with HST~\cite{Todarello:2024qci}. The figure also includes bounds on deviations from Newtonian gravity~\cite{Chen:2014oda}, parametrized by the Yukawa potential in Eq.~\eqref{eq:yukawa}. For convenience, the upper axis shows the corresponding interaction range for each mass, while the right axis indicates the effective Planck mass associated with the spin-2 field.}
\label{fig:results}
\end{figure}

\subsection{Solar Dark Radiation Bound}

The total power emitted from the Sun in the form of spin-2 particles is obtained by integrating the energy emission rate per volume  
\begin{equation}
L_\odot \Big|_{\delta M} = \int_{\rm Sun} dV \int \mathrm{d}\omega\, \omega\, \frac{\mathrm{d}\Gamma_{\rm tot}}{\mathrm{d}\omega\, \mathrm{d}V}\,,
\label{eq:psun}
\end{equation}
with ${\mathrm{d}\Gamma_{\rm tot}}/{\mathrm{d}\omega\, \mathrm{d}V}$  given in Eq.~\eqref{eq:GammaMassive} adding over all channels. Numerically, for sub-keV mass particles we find
\begin{equation}
L_\odot \Big|_{\delta M} \simeq  \left(1.5 \times 10^{15}\ {\rm erg\, s^{-1}}\right) \times \frac{G'}{G}\,.
\end{equation}

Observations of the all-flavor solar neutrino flux indicate that nuclear burning accounts for the Sun’s total luminosity within about 10\%, leaving only a small fraction available for any exotic energy loss~\cite{Gondolo:2008dd}. We therefore impose the conservative requirement
\begin{equation}
L_\odot \Big|_{\delta M}  \le 0.1 L_\odot \,,
\end{equation}
which ensures that spin-2 emission does not exceed this residual energy budget. This directly limits the coupling as\footnote{Eq.~\eqref{eq:solarBound} differs numerically from those reported in Ref.~\cite{Cembranos:2017vgi}. 
This discrepancy most likely originates from their omission of plasma screening effects. } 
\begin{align}
G'< 2.5 \times 10^{17} G, && \text{for}\quad m \lesssim 1~\text{keV} &&\text{[Solar dark radiation bound]}\,.
\label{eq:solarBound}
\end{align}
The resulting constraint is shown in Fig.~\ref{fig:results}. 

\subsection{Horizontal-Branch Stars}

More stringent limits arise from observations of horizontal-branch (HB) stars in globular clusters, whose lifetimes are sensitive to any additional cooling channels beyond standard helium burning. The emission of spin-2 particles would enhance the stellar energy loss, thereby reducing the duration of the HB phase if the coupling were sufficiently large.

During the Helium-burning stage, HB stars are characterized by typical core conditions of $T \simeq 10^{8}~{\rm K} =8.6~{\rm keV} $ and $\rho \simeq 0.6\times10^{4}~{\rm g\,cm^{-3}}$. In this regime, the plasma  is non-degenerate, which allows the use of the same formalism developed for the Sun to compute the relevant emission processes. In this manner, the requirement that the HB lifetime agrees with observations imposes the bound~\cite{Raffelt:1996wa} 
\begin{align}
\epsilon \big|_{\delta M} \lesssim \, 10\, {\rm erg\,g^{-1}\,s^{-1}}\,,
&&
\text{where}
&&
\epsilon \big|_{\delta M} =\frac{1}{\rho} \int \mathrm{d}\omega\, \omega\, \frac{\mathrm{d}\Gamma_{\rm tot}}{\mathrm{d}\omega\, \mathrm{d}V}\,,
\end{align}
which translates to
\begin{align}
G' &< 1.3 \times 10^{15} G, && \text{for}\quad m \lesssim 8~\text{keV} && \text{[HB bound]}
\label{eq:HBbound}
\end{align}

More sophisticated analyses, such as those carried out for axion emission in Ref.~\cite{ayala:2014pea}, which incorporate detailed stellar evolution modeling, are expected to yield limits of comparable magnitude. 

\subsection{Helioscopes and Photon Conversion of Spin-2 Particles in Magnetic Fields}
\label{sec:helioscopes}

As discussed in Sec.~\ref{sec:photoconversion}, spin-2 particles can convert into photons through a mechanism closely analogous to axion–photon mixing. 
We now exploit this correspondence to reinterpret existing axion helioscope searches as probes of light spin-2 particles. 
This recasting combines the photon–spin-2 conversion formalism introduced earlier with the detailed solar emission spectra derived in Sec.~\ref{sec:emission}, particularly the results shown in both panels of Fig.~\ref{fig:spectrum}. 
By comparing Eq.~\eqref{eq:PrimakoffConversion} with Eq.~\eqref{eq:conversion0}, and the flux in Fig.~\ref{fig:spectrum} with the standard Primakoff axion flux (see e.g.~\cite{CAST:2017uph}), limits on the axion–photon coupling can be translated into bounds on the coupling ratio $G'/G$. Let us discuss each experiment separately. 

\begin{itemize}

\item \textbf{CAST.}  
As mentioned above, CAST searches for solar axions via their conversion into X-ray photons in a strong magnetic field. It employs a $9$~T, $9.3$~m long LHC dipole magnet with low-background X-ray detectors aligned with the Sun~\cite{CAST:2017uph}. 
Its well-characterized magnetic configuration also makes it sensitive to light spin-2 particles that convert into photons through the interaction in Eq.~\eqref{eq:Lmix}. The resulting constraint is shown in Fig.~\ref{fig:results}. 

\item \textbf{Prospects for Spin-2 Searches at BabyIAXO and IAXO.}  
The International Axion Observatory (IAXO) and its intermediate prototype, BabyIAXO, are next-generation helioscopes designed to detect solar axions through their conversion into photons in a strong magnetic field~\cite{IAXO:2025ltd}. Their large magnetic volume, high field strength, and ultra-low background detectors make them sensitive to light bosons beyond axions, including spin-2 particles. Using the conversion rates in Eq.~\eqref{eq:conversion0} with the appropriate $B$, $L$, from Ref.~\cite{IAXO:2020wwp}  we compute the projected reach in $G'/G$. BabyIAXO, expected to begin operation in the late 2020s, will probe couplings near the HB bound, while the full IAXO experiment will extend this reach by more than an order of magnitude. The resulting constraints are shown in Fig.~\ref{fig:results}. 

\item \textbf{High Masses.}  
At small masses, the conversion of solar spin-2 particles into photons inside the magnet proceeds coherently, maximizing $P_{\delta M\to\gamma}$ when $qL\ll1$. As the mass increases, coherence is lost, leading to suppression around $m\sim0.1~\text{eV}$ (see Fig.~\ref{fig:results}). To restore sensitivity, photons must acquire an effective mass comparable to the spin-2 particle mass. This can be achieved by filling the magnet bores with a low-$Z$ buffer gas, such as helium, which gives photons a plasma mass that compensates the momentum mismatch. By tuning the gas density, one can match the photon mass to different hypothetical spin-2 masses. Following Ref.~\cite{Galan:2015msa}, we show the resulting extrapolations for BabyIAXO and IAXO in Fig.~\ref{fig:results} for $m>0.1~\text{eV}$.
\end{itemize}

Nevertheless, as will be discussed below, the parameter space accessible to helioscopes largely overlaps with regions already excluded by inverse-square law tests, implying that they will not probe untested parameter space for light spin-2 particles.

\subsection{Constraints from Tests of the Inverse-Square Law}
\label{sec:fifthforce}
As shown in Eq.~\eqref{eq:yukawa}, the exchange of a massive spin-2 particle between two non-relativistic bodies produces a Yukawa-type correction to the Newtonian potential. Such deviations from the inverse-square law are subject to extremely stringent experimental limits, obtained from high-precision measurements of short-range gravitational forces. These constraints place some of the strongest existing bounds on new light mediators, including massive spin-2 particles. Comprehensive reviews of these experiments and their implications for new light-force mediators can be found in Refs.~\cite{Adelberger:2003zx, Adelberger:2009zz}. 

For the mass range relevant to this work, the most constraining results are those of Ref.~\cite{Chen:2014oda}, which tested the gravitational inverse-square law at submicron distances using a ``Casimir-less'' setup~\cite{Decca:2005qz}. The experiment measured the differential force between a gold-coated test mass mounted on a microelectromechanical torsional oscillator and a rotating source mass made of alternating gold and silicon sectors. By covering the source with a sufficiently thick gold layer, the dominant Casimir background was suppressed, isolating any possible mass-dependent Yukawa contribution.

Using lock-in detection at the resonance frequency of the oscillator, the authors achieved sub-femto-Newton sensitivity and derived upper limits on deviations from Newtonian gravity over interaction ranges of $10$--$2000~\mathrm{nm}$. For the present study, these results translate directly into very strong constraints on the coupling ratio $G'/G$, which are shown in Fig.~\ref{fig:results}.

To ensure that these constraints apply within the linear regime of the theory, the Vainshtein radius associated with the test masses must remain smaller than the experimental separation scale. 
For two spherical masses separated by a distance $d$, the generalized Vainshtein radius is given by Eq.~\eqref{eq:rV}. 
Imposing $r_{\text{V}} < d$ guarantees the validity of the Yukawa potential in Eq.~\eqref{eq:yukawa}, leading to
\begin{equation}
m \gtrsim 5\times 10^{-5}~{\rm eV}\,
\left(\frac{G'/G}{10^{15}}\right)^{1/2}
\left(\frac{M}{50~\mu{\rm g}}\right)^{1/2}
\left(\frac{d}{10~{\rm nm}}\right)^{-3/2},
\end{equation}
valid for $G'\!\gg\!G$.
Hence, the assumed range of validity of bimetric theories encompasses the parameter space shown in Fig.~\ref{fig:results}.

\subsection{Indirect Constraints
from Photon Searches}

At higher masses the decay of the spin-2 dark matter field into two photons can be constrained through indirect searches for narrow astrophysical spectral lines. 
These observations provide complementary limits  on the  coupling ratio $G'/G$, as shown in Fig.~\ref{fig:results}. 
To obtain these bounds, we recast existing limits---typically expressed in terms of the axion--photon coupling $g_{a\gamma\gamma}$---into constraints on $G'/G$ by comparing Eq.~\eqref{eq:Gamma_a} with Eq.~\eqref{eq:GammaDeltaM}.

\begin{itemize}
\item {\bf Bounds from Hubble Space Telescope (HST) Ultraviolet Spectroscopy.} The HST analysis of Ref.~\cite{Todarello:2024qci} searched for ultraviolet emission lines from dark-matter-dominated systems—specifically the Draco and Ursa Minor dwarf spheroidals and the Virgo and Fornax clusters. The expected signal from particle decay is a narrow line centered at half the particle mass. Using data for wavelengths between 110 and 170~nm, the authors found no excess consistent with a decay signal. The absence of such features constrains the decay rate of dark matter into two photons in the corresponding mass range (here, $m \sim 15$–$20$~eV). 
\item {\bf Heating of the Leo~T Dwarf Galaxy.} The Leo~T dwarf galaxy analysis of Ref.~\cite{Wadekar:2021qae} provides an independent and highly complementary constraint. This galaxy contains cold neutral hydrogen gas with extremely low radiative cooling, making it very sensitive to any additional heating from dark-matter decay or annihilation. By requiring that the energy deposited by decay-induced photons not exceed the measured cooling rate, the study places strong bounds on $G'/G$ for photon energies between about $10$~eV and $1$~keV.
\end{itemize}

\subsection{Discussion}

As shown in Fig.~\ref{fig:results}, inverse-square law tests probe a significant portion of the parameter space, excluding the region below $\mathcal{O}(1~\text{eV})$ that could otherwise be accessed by helioscope experiments. This exclusion arises because modifications of the gravitational force at distances larger than a few micrometers are strongly ruled out by precision laboratory measurements. Unlike axions, spin-2 particles mediate long-range interactions and thus induce observable deviations from Newtonian gravity. Axions, being pseudoscalars, do not produce such effects, which explains why their allowed parameter space remains considerably broader at low masses.

At higher masses, indirect searches for dark matter decaying into photons constrain complementary regions of parameter space, typically above $\sim 20~\text{eV}$. These bounds are particularly strong in the X-ray range, where decay photons can heat the interstellar gas in dark matter–dominated systems. The most stringent limits arise from the non-observation of excess heating in the Leo~T dwarf galaxy~\cite{ Wadekar:2021qae}. 

The limits from dark-matter decay and from inverse-square law tests overlap around $m\!\sim\!10~\text{eV}$, both being comparable in strength to the  solar energy-loss bound derived in this work.
In contrast, the constraint derived from HB stars proves to be extremely competitive. Under the well-established criterion that anomalous energy losses in HB cores should not exceed approximately $10~\text{erg\,g}^{-1}\text{s}^{-1}$~\cite{Raffelt:1996wa}, our calculations yield limits on $G'/G$ that rival those from inverse-square law experiments and dark matter decay searches, especially for masses between $\sim3$ and $20~\text{eV}$. This constitutes the main result of the present study: HB stars provide one of the most stringent and robust constraints on the coupling of light spin-2 particles to matter.

Our results are fairly general and do not depend on the specific details of bimetric models. Barring the problem of the Boulware ghosts, our analysis can be adapted to weakly coupled, propagating spin-2 field. Nevertheless, bimetric theories provide a useful theoretical framework, as they offer a ghost-free realization of massive spin-2 particles. It is worth noting that our limits cannot be directly applied to Kaluza–Klein gravitons, since those form discrete towers of modes whose cumulative phenomenology differs from that of a single isolated state, see cf. Refs.~\cite{Arkani-Hamed:1998sfv,Hannestad:2001jv,Hannestad:2001xi,Hannestad:2003yd,Hardy:2025ajb}.

Finally, the methods developed here can be extended to degenerate stellar environments. As discussed in Ref.~\cite{Raffelt:1996wa}, dense objects such as neutron stars, white dwarfs, or red giants may impose similar or even stronger limits, provided the emission rates are adapted to account for degeneracy. 
Incorporating these effects into the calculation of spin-2 emission will be the subject of future work, potentially leading to even stronger astrophysical constraints on light spin-2 interactions.

\section{Conclusions}
\label{sec:conclusions}

We have presented a comprehensive analysis of stellar and laboratory constraints on light spin-2 particles, motivated by bimetric gravity and related theories featuring a weakly coupled massive tensor field. Using a consistent treatment of the solar and stellar plasma, including screening effects, we have computed the production rates of spin-2 particles through bremsstrahlung and photoproduction processes and derived their corresponding emission spectra. The result is shown in Fig.~\ref{fig:spectrum}.

By comparing the predicted stellar energy losses with observations, we derived limits on the effective gravitational coupling ratio $G'/G$ across a broad mass range, from sub-eV to a few keV. These bounds complement existing laboratory, astrophysical, and cosmological constraints. The strongest limits arise from HB stars, which exclude coupling strengths comparable to or stronger than those from inverse-square law tests and dark matter decay searches.
Solar energy-loss arguments yield weaker bounds at intermediate masses. At higher masses ($m \gtrsim 20$ eV), indirect limits from photon-based searches, particularly heating constraints from the Leo~T dwarf galaxy, further restrict the remaining parameter space.

Our results are general and apply to any weakly interacting, propagating spin-2 field, independently of the specific realization within bimetric gravity. They cannot be directly translated to models involving Kaluza–Klein gravitons, whose collective behavior differs from that of a single light tensor state. Finally, the formalism developed here can be extended to degenerate stellar environments such as white dwarfs, neutron stars, and red giants. Work along these lines will be presented in future studies.

{\section*{Acknowledgements}
%
We thank Arturo De Giorgi, Javier Gal\'an, Maurizio Giannotti, Igor Irastorza, Giuseppe Lucente, Alessandro Mirizzi and Nuria Rius for valuable discussions. 
A.R. acknowledges support by the Deutsche Forschungsgemeinschaft (DFG, German Research Foundation) under Germany’s Excellence Strategy - EXC 2121 Quantum Universe - 390833306.  CGC is supported by a Ramón y Cajal contract with Ref.~RYC2020-029248-I, the Spanish National Grant PID2022-137268NA-C55 and Generalitat Valenciana through the grant CIPROM/22/69. This work has been partially funded by the Deutsche Forschungsgemeinschaft (DFG, German Research Foundation) - 491245950.
This article is based upon work from COST Action COSMIC WISPers CA21106, supported by COST (European Cooperation in Science and Technology). 

\appendix

\section{Bremsstrahlung Cross Sections}
\label{sec:appA}

Table~\ref{table:processes} summarizes the bremsstrahlung cross sections to leading order in the screening parameter. 
In this appendix, we present the complete expressions. 
Recall that
\begin{align}
L = \log \sqrt{\frac{(1+\xi)^2+\xi _s^2}{(1-\xi )^2+\xi _s^2}}, \qquad \text{where} \qquad \xi_s = \frac{\kappa}{p_i}\,.
\end{align}

\paragraph{Tensor modes ($\lambda=\pm2$).}  We find~\cite{Garcia-Cely:2024ujr}
\begin{eqnarray}
\label{eq:apeZ}
\!\!\!\!\frac{{\mathrm d}\sigma v}{{\mathrm d}\omega}
\Bigg|_{eZ\to eZ \delta M}
\!\!\!\!\!\!\!\!\!\!\!\!&=&
\frac{32\, Z^2 \alpha^2 G\, p_i}{15\, \omega}
\left(\frac{1}{m_e} + \frac{1}{m_Z}\right)
\Bigg[
3(1+\xi^2+\xi_s^2)L
+10\xi \nonumber\\
&&
+\frac{1}{3}\xi_s^2
\left(
\frac{(1-\xi)^2
\left[
18(1+\xi)^4 + 29(1+\xi)^2\xi_s^2 + 12\xi_s^4
\right]}
{\left((1+\xi)^2+\xi_s^2\right)^3}
-(\xi \to -\xi)
\right)
\Bigg].
\end{eqnarray}
We do not assume $m_Z \gg m_e$, but simply that both nucleons and electrons are non-relativistic.  

\begin{eqnarray}
\label{eq:apee}
\!\!\!\!\frac{{\mathrm d}\sigma v}{{\mathrm d}\omega}
\Bigg|_{ee\to ee\, \delta M}
\!\!\!\!\!\!\!\!\!\!\!\!&=&
\frac{32\, \alpha^2 G\, p_i}{15\, \omega\, m_e}
\Bigg\{
20\xi
-\frac{6\xi(1+\xi^4)}{(1+\xi^2)^2}
+\Bigg[
6(1+\xi^2)
-\frac{3(1-\xi^2)^4 + 7(1-\xi^4)^2}{2(1+\xi^2)^3} \nonumber\\
&&
+\frac{1}{2}\xi_s^2
\Bigg(
\frac{6(\xi^4+1)(1-\xi^2)^2}
{(\xi^2+1)^2(\xi^2+\xi_s^2+1)^2}
+\frac{2(\xi^4-4\xi^2+1)(1-\xi^2)^2}
{(\xi^2+1)(\xi^2+\xi_s^2+1)^3}
\nonumber\\
&&
+\frac{13\xi^8 + 22\xi^4 + 13}
{(\xi^2+1)^3(\xi^2+\xi_s^2+1)}
+15
\Bigg)
\Bigg]L
+\xi_s^2
\Bigg[
2(1-\xi^2)^2
\Bigg(
-\frac{(1-\xi)^4 - 80\xi^2}
{16\,\xi^2(\xi+1)^2((\xi+1)^2+\xi_s^2)}
\nonumber\\
&&
-\frac{3(\xi^4+1)\xi_s^2 + 4(\xi^6+1)}
{8\xi(\xi^2+1)^2(\xi^2+\xi_s^2+1)^2}
+\frac{6(\xi+1)^2 + 5\xi_s^2}
{3((\xi+1)^2+\xi_s^2)^3}
\Bigg)
-(\xi \to -\xi)
\Bigg]
\Bigg\}.
\end{eqnarray}
They agree with the cross sections for the bremsstrahlung emission of ordinary gravitons of the same helicity.
For $\xi_s = 0$, these cross sections reproduce those of 
Refs.~\cite{Gould:1981an, Gould:1985}.

\paragraph{Scalar mode ($\lambda=0$).}

\begin{align}
\!\!\!\!\!\!\!\!\frac{{\mathrm d}\sigma v}{{\mathrm d}\omega}
\Bigg|_{ee\to ee\, \delta M}
&=
\frac{16\, Z^2\alpha^2\,G\,  p_i}{45\, \omega}
\left(\frac{1}{m_e} + \frac{1}{m_Z}\right)
\Bigg[
 (1+\xi^2+\xi_s^{2})\,L
 +30\,\xi \nonumber\\[4pt]
&\quad
+ \frac{2}{3}\,\xi_s^{2}
\left(
\frac{(1-\xi)^2
\left[
18(1+\xi)^4 + 29(1+\xi)^2\xi_s^2 + 12\xi_s^4
\right]}
{\left((1+\xi)^2+\xi_s^2\right)^3}
- (\xi \to -\xi)
\right)
\Bigg].
\label{eq:ee_scattering}
\end{align}

\begin{eqnarray}
\label{eq:apeeSPIN0}
\!\!\!\!\frac{{\mathrm d}\sigma v}{{\mathrm d}\omega}
\Bigg|_{ee\to ee\, \delta M}
\!\!\!\!\!\!\!\!\!\!\!\!&=&
\frac{32\, \alpha^2 G\, p_i}{15\, \omega\, m_e}
\Bigg\{
\frac{29}{3}\xi
+\frac{2\xi^3}{3(1+\xi^2)^2}
+\Bigg[
\frac{1}{3}(1+\xi^2)
-\frac{(1-\xi^2)^4 + 29(1-\xi^4)^2}{12(1+\xi^2)^3} \\
&&
+\frac{1}{2}\xi_s^2
\Bigg(
\frac{2(3\xi^4+5\xi^2+3)(1-\xi^2)^2}
{3(\xi^2+1)^2(\xi^2+\xi_s^2+1)^2}
+\frac{2(\xi^4+\xi^2+1)(1-\xi^2)^2}
{3(\xi^2+1)(\xi^2+\xi_s^2+1)^3}
\nonumber\\
&&
+\frac{31\xi^8 - 46\xi^4 + 31}
{6(\xi^2+1)^3(\xi^2+\xi_s^2+1)}
+\frac{5}{6}
\Bigg)
\Bigg]L
+\xi_s^2
\Bigg[
\frac{2}{3}(1-\xi^2)^2
\Bigg(
-\frac{(1-\xi)^4 - 80\xi^2}
{16\,\xi^2(\xi+1)^2((\xi+1)^2+\xi_s^2)}
\nonumber\\
&&
-\frac{(3+5\xi^2+3\xi^4)\xi_s^2 + 2(1+\xi^2)(2+3\xi^2+2\xi^4)}
{8\xi(\xi^2+1)^2(\xi^2+\xi_s^2+1)^2}
+\frac{6(\xi+1)^2 + 5\xi_s^2}
{3((\xi+1)^2+\xi_s^2)^3}
\Bigg)
-(\xi \to -\xi)
\Bigg]
\Bigg\}.\nonumber
\end{eqnarray}

\bibliographystyle{JHEP}
\bibliography{ref}

\providecommand{\href}[2]{#2}\begingroup\raggedright\begin{thebibliography}{10}

\bibitem{Hassan:2011hr}
S.F.~Hassan and R.A.~Rosen, \emph{{Resolving the Ghost Problem in non-Linear
  Massive Gravity}},
  \href{https://doi.org/10.1103/PhysRevLett.108.041101}{\emph{Phys. Rev. Lett.}
  {\bfseries 108} (2012) 041101}
  [\href{https://arxiv.org/abs/1106.3344}{{\ttfamily 1106.3344}}].

\bibitem{Hassan:2011tf}
S.F.~Hassan, R.A.~Rosen and A.~Schmidt-May, \emph{{Ghost-free Massive Gravity
  with a General Reference Metric}},
  \href{https://doi.org/10.1007/JHEP02(2012)026}{\emph{JHEP} {\bfseries 02}
  (2012) 026} [\href{https://arxiv.org/abs/1109.3230}{{\ttfamily 1109.3230}}].

\bibitem{Hassan:2012wr}
S.F.~Hassan, A.~Schmidt-May and M.~von Strauss, \emph{{On Consistent Theories
  of Massive Spin-2 Fields Coupled to Gravity}},
  \href{https://doi.org/10.1007/JHEP05(2013)086}{\emph{JHEP} {\bfseries 05}
  (2013) 086} [\href{https://arxiv.org/abs/1208.1515}{{\ttfamily 1208.1515}}].

\bibitem{deRham:2010kj}
C.~de~Rham, G.~Gabadadze and A.J.~Tolley, \emph{{Resummation of Massive
  Gravity}}, \href{https://doi.org/10.1103/PhysRevLett.106.231101}{\emph{Phys.
  Rev. Lett.} {\bfseries 106} (2011) 231101}
  [\href{https://arxiv.org/abs/1011.1232}{{\ttfamily 1011.1232}}].

\bibitem{Schmidt-May:2015vnx}
A.~Schmidt-May and M.~von Strauss, \emph{{Recent developments in bimetric
  theory}}, \href{https://doi.org/10.1088/1751-8113/49/18/183001}{\emph{J.
  Phys. A} {\bfseries 49} (2016) 183001}
  [\href{https://arxiv.org/abs/1512.00021}{{\ttfamily 1512.00021}}].

\bibitem{deRham:2014zqa}
C.~de~Rham, \emph{{Massive Gravity}},
  \href{https://doi.org/10.12942/lrr-2014-7}{\emph{Living Rev. Rel.} {\bfseries
  17} (2014) 7} [\href{https://arxiv.org/abs/1401.4173}{{\ttfamily
  1401.4173}}].

\bibitem{Maeda:2013bha}
K.-i.~Maeda and M.S.~Volkov, \emph{{Anisotropic universes in the ghost-free
  bigravity}}, \href{https://doi.org/10.1103/PhysRevD.87.104009}{\emph{Phys.
  Rev. D} {\bfseries 87} (2013) 104009}
  [\href{https://arxiv.org/abs/1302.6198}{{\ttfamily 1302.6198}}].

\bibitem{Aoki:2016zgp}
K.~Aoki and S.~Mukohyama, \emph{{Massive gravitons as dark matter and
  gravitational waves}},
  \href{https://doi.org/10.1103/PhysRevD.94.024001}{\emph{Phys. Rev. D}
  {\bfseries 94} (2016) 024001}
  [\href{https://arxiv.org/abs/1604.06704}{{\ttfamily 1604.06704}}].

\bibitem{Babichev:2016hir}
E.~Babichev, L.~Marzola, M.~Raidal, A.~Schmidt-May, F.~Urban, H.~Veerm{\"a}e
  et~al., \emph{{Bigravitational origin of dark matter}},
  \href{https://doi.org/10.1103/PhysRevD.94.084055}{\emph{Phys. Rev. D}
  {\bfseries 94} (2016) 084055}
  [\href{https://arxiv.org/abs/1604.08564}{{\ttfamily 1604.08564}}].

\bibitem{Babichev:2016bxi}
E.~Babichev, L.~Marzola, M.~Raidal, A.~Schmidt-May, F.~Urban, H.~Veerm{\"a}e
  et~al., \emph{{Heavy spin-2 Dark Matter}},
  \href{https://doi.org/10.1088/1475-7516/2016/09/016}{\emph{JCAP} {\bfseries
  09} (2016) 016} [\href{https://arxiv.org/abs/1607.03497}{{\ttfamily
  1607.03497}}].

\bibitem{Dubovsky:2004ud}
S.L.~Dubovsky, P.G.~Tinyakov and I.I.~Tkachev, \emph{{Massive graviton as a
  testable cold dark matter candidate}},
  \href{https://doi.org/10.1103/PhysRevLett.94.181102}{\emph{Phys. Rev. Lett.}
  {\bfseries 94} (2005) 181102}
  [\href{https://arxiv.org/abs/hep-th/0411158}{{\ttfamily hep-th/0411158}}].

\bibitem{Marzola:2017lbt}
L.~Marzola, M.~Raidal and F.R.~Urban, \emph{{Oscillating Spin-2 Dark Matter}},
  \href{https://doi.org/10.1103/PhysRevD.97.024010}{\emph{Phys. Rev. D}
  {\bfseries 97} (2018) 024010}
  [\href{https://arxiv.org/abs/1708.04253}{{\ttfamily 1708.04253}}].

\bibitem{Chu:2017msm}
X.~Chu and C.~Garcia-Cely, \emph{{Self-interacting Spin-2 Dark Matter}},
  \href{https://doi.org/10.1103/PhysRevD.96.103519}{\emph{Phys. Rev. D}
  {\bfseries 96} (2017) 103519}
  [\href{https://arxiv.org/abs/1708.06764}{{\ttfamily 1708.06764}}].

\bibitem{Cai:2021nmk}
H.~Cai, G.~Cacciapaglia and S.J.~Lee, \emph{{Massive Gravitons as Feebly
  Interacting Dark Matter Candidates}},
  \href{https://doi.org/10.1103/PhysRevLett.128.081806}{\emph{Phys. Rev. Lett.}
  {\bfseries 128} (2022) 081806}
  [\href{https://arxiv.org/abs/2107.14548}{{\ttfamily 2107.14548}}].

\bibitem{Kolb:2023dzp}
E.W.~Kolb, S.~Ling, A.J.~Long and R.A.~Rosen, \emph{{Cosmological gravitational
  particle production of massive spin-2 particles}},
  \href{https://doi.org/10.1007/JHEP05(2023)181}{\emph{JHEP} {\bfseries 05}
  (2023) 181} [\href{https://arxiv.org/abs/2302.04390}{{\ttfamily
  2302.04390}}].

\bibitem{Cembranos:2017vgi}
J.A.R.~Cembranos, A.L.~Maroto and H.~Villarrubia-Rojo, \emph{{Constraints on
  hidden gravitons from fifth-force experiments and stellar energy loss}},
  \href{https://doi.org/10.1007/JHEP09(2017)104}{\emph{JHEP} {\bfseries 09}
  (2017) 104} [\href{https://arxiv.org/abs/1706.07818}{{\ttfamily
  1706.07818}}].

\bibitem{Garcia-Cely:2024ujr}
C.~Garc{\'\i}a-Cely and A.~Ringwald, \emph{{Complete Gravitational-Wave
  Spectrum of the Sun}}, \href{https://doi.org/10.1103/gtwg-pr41}{\emph{Phys.
  Rev. Lett.} {\bfseries 135} (2025) 061001}
  [\href{https://arxiv.org/abs/2407.18297}{{\ttfamily 2407.18297}}].

\bibitem{Raffelt:1996wa}
G.G.~Raffelt, \emph{{Stars as laboratories for fundamental physics}: {The
  astrophysics of neutrinos, axions, and other weakly interacting particles}}
  (5, 1996).

\bibitem{Fierz:1939ix}
M.~Fierz and W.~Pauli, \emph{{On relativistic wave equations for particles of
  arbitrary spin in an electromagnetic field}},
  \href{https://doi.org/10.1098/rspa.1939.0140}{\emph{Proc. Roy. Soc. Lond. A}
  {\bfseries 173} (1939) 211}.

\bibitem{Boulware:1972yco}
D.G.~Boulware and S.~Deser, \emph{{Can gravitation have a finite range?}},
  \href{https://doi.org/10.1103/PhysRevD.6.3368}{\emph{Phys. Rev. D} {\bfseries
  6} (1972) 3368}.

\bibitem{Hagiwara:2008jb}
K.~Hagiwara, J.~Kanzaki, Q.~Li and K.~Mawatari, \emph{{HELAS and
  MadGraph/MadEvent with spin-2 particles}},
  \href{https://doi.org/10.1140/epjc/s10052-008-0663-x}{\emph{Eur. Phys. J. C}
  {\bfseries 56} (2008) 435} [\href{https://arxiv.org/abs/0805.2554}{{\ttfamily
  0805.2554}}].

\bibitem{Christensen:2013aua}
N.D.~Christensen, P.~de~Aquino, N.~Deutschmann, C.~Duhr, B.~Fuks,
  C.~Garcia-Cely et~al., \emph{{Simulating spin-$ \frac{3}{2}$ particles at
  colliders}}, \href{https://doi.org/10.1140/epjc/s10052-013-2580-x}{\emph{Eur.
  Phys. J. C} {\bfseries 73} (2013) 2580}
  [\href{https://arxiv.org/abs/1308.1668}{{\ttfamily 1308.1668}}].

\bibitem{deRham:2012ew}
C.~de~Rham, G.~Gabadadze, L.~Heisenberg and D.~Pirtskhalava,
  \emph{{Nonrenormalization and naturalness in a class of scalar-tensor
  theories}}, \href{https://doi.org/10.1103/PhysRevD.87.085017}{\emph{Phys.
  Rev. D} {\bfseries 87} (2013) 085017}
  [\href{https://arxiv.org/abs/1212.4128}{{\ttfamily 1212.4128}}].

\bibitem{Alloul:2013bka}
A.~Alloul, N.D.~Christensen, C.~Degrande, C.~Duhr and B.~Fuks, \emph{{FeynRules
  2.0 - A complete toolbox for tree-level phenomenology}},
  \href{https://doi.org/10.1016/j.cpc.2014.04.012}{\emph{Comput. Phys. Commun.}
  {\bfseries 185} (2014) 2250}
  [\href{https://arxiv.org/abs/1310.1921}{{\ttfamily 1310.1921}}].

\bibitem{Belyaev:2012qa}
A.~Belyaev, N.D.~Christensen and A.~Pukhov, \emph{{CalcHEP 3.4 for collider
  physics within and beyond the Standard Model}},
  \href{https://doi.org/10.1016/j.cpc.2013.01.014}{\emph{Comput. Phys. Commun.}
  {\bfseries 184} (2013) 1729}
  [\href{https://arxiv.org/abs/1207.6082}{{\ttfamily 1207.6082}}].

\bibitem{Patel:2016fam}
H.H.~Patel, \emph{{Package-X 2.0: A Mathematica package for the analytic
  calculation of one-loop integrals}},
  \href{https://doi.org/10.1016/j.cpc.2017.04.015}{\emph{Comput. Phys. Commun.}
  {\bfseries 218} (2017) 66}
  [\href{https://arxiv.org/abs/1612.00009}{{\ttfamily 1612.00009}}].

\bibitem{Choi:1994ax}
S.Y.~Choi, J.S.~Shim and H.S.~Song, \emph{{Factorization and polarization in
  linearized gravity}},
  \href{https://doi.org/10.1103/PhysRevD.51.2751}{\emph{Phys. Rev. D}
  {\bfseries 51} (1995) 2751}
  [\href{https://arxiv.org/abs/hep-th/9411092}{{\ttfamily hep-th/9411092}}].

\bibitem{Donoghue:1994dn}
J.F.~Donoghue, \emph{{General relativity as an effective field theory: The
  leading quantum corrections}},
  \href{https://doi.org/10.1103/PhysRevD.50.3874}{\emph{Phys. Rev. D}
  {\bfseries 50} (1994) 3874}
  [\href{https://arxiv.org/abs/gr-qc/9405057}{{\ttfamily gr-qc/9405057}}].

\bibitem{Han:1998sg}
T.~Han, J.D.~Lykken and R.-J.~Zhang, \emph{{On Kaluza-Klein states from large
  extra dimensions}},
  \href{https://doi.org/10.1103/PhysRevD.59.105006}{\emph{Phys. Rev. D}
  {\bfseries 59} (1999) 105006}
  [\href{https://arxiv.org/abs/hep-ph/9811350}{{\ttfamily hep-ph/9811350}}].

\bibitem{Strumia:2025dfn}
A.~Strumia and G.~Landini, \emph{{Optical gravitational waves as signals of
  gravitationally-decaying particles}},
  \href{https://doi.org/10.1007/JHEP04(2025)068}{\emph{JHEP} {\bfseries 04}
  (2025) 068} [\href{https://arxiv.org/abs/2501.09794}{{\ttfamily
  2501.09794}}].

\bibitem{Dunsky:2025pvd}
D.I.~Dunsky, G.~Krnjaic and E.~Pinetti, \emph{{Observing Dark Matter Decays to
  Gravitons via Graviton-Photon Conversion}},
  \href{https://arxiv.org/abs/2503.19019}{{\ttfamily 2503.19019}}.

\bibitem{Cembranos:2025uhe}
J.A.R.~Cembranos and {\'A}.~Cendal, \emph{{Sensitivity forecasts for
  gravitational-wave detectors to dark matter decaying into gravitons}},
  \href{https://arxiv.org/abs/2510.18958}{{\ttfamily 2510.18958}}.

\bibitem{Peccei:1977hh}
R.D.~Peccei and H.R.~Quinn, \emph{{CP Conservation in the Presence of
  Instantons}}, \href{https://doi.org/10.1103/PhysRevLett.38.1440}{\emph{Phys.
  Rev. Lett.} {\bfseries 38} (1977) 1440}.

\bibitem{Weinberg:1977ma}
S.~Weinberg, \emph{{A New Light Boson?}},
  \href{https://doi.org/10.1103/PhysRevLett.40.223}{\emph{Phys. Rev. Lett.}
  {\bfseries 40} (1978) 223}.

\bibitem{Wilczek:1977pj}
F.~Wilczek, \emph{{Problem of Strong $P$ and $T$ Invariance in the Presence of
  Instantons}}, \href{https://doi.org/10.1103/PhysRevLett.40.279}{\emph{Phys.
  Rev. Lett.} {\bfseries 40} (1978) 279}.

\bibitem{Vafa:1984xg}
C.~Vafa and E.~Witten, \emph{{Parity Conservation in QCD}},
  \href{https://doi.org/10.1103/PhysRevLett.53.535}{\emph{Phys. Rev. Lett.}
  {\bfseries 53} (1984) 535}.

\bibitem{Preskill:1982cy}
J.~Preskill, M.B.~Wise and F.~Wilczek, \emph{{Cosmology of the Invisible
  Axion}}, \href{https://doi.org/10.1016/0370-2693(83)90637-8}{\emph{Phys.
  Lett. B} {\bfseries 120} (1983) 127}.

\bibitem{Abbott:1982af}
L.F.~Abbott and P.~Sikivie, \emph{{A Cosmological Bound on the Invisible
  Axion}}, \href{https://doi.org/10.1016/0370-2693(83)90638-X}{\emph{Phys.
  Lett. B} {\bfseries 120} (1983) 133}.

\bibitem{Dine:1982ah}
M.~Dine and W.~Fischler, \emph{{The Not So Harmless Axion}},
  \href{https://doi.org/10.1016/0370-2693(83)90639-1}{\emph{Phys. Lett. B}
  {\bfseries 120} (1983) 137}.

\bibitem{Arias:2012az}
P.~Arias, D.~Cadamuro, M.~Goodsell, J.~Jaeckel, J.~Redondo and A.~Ringwald,
  \emph{{WISPy Cold Dark Matter}},
  \href{https://doi.org/10.1088/1475-7516/2012/06/013}{\emph{JCAP} {\bfseries
  06} (2012) 013} [\href{https://arxiv.org/abs/1201.5902}{{\ttfamily
  1201.5902}}].

\bibitem{Primakoff:1951iae}
H.~Primakoff, \emph{{Photoproduction of neutral mesons in nuclear electric
  fields and the mean life of the neutral meson}},
  \href{https://doi.org/10.1103/PhysRev.81.899}{\emph{Phys. Rev.} {\bfseries
  81} (1951) 899}.

\bibitem{Dicus:1978fp}
D.A.~Dicus, E.W.~Kolb, V.L.~Teplitz and R.V.~Wagoner, \emph{{Astrophysical
  Bounds on the Masses of Axions and Higgs Particles}},
  \href{https://doi.org/10.1103/PhysRevD.18.1829}{\emph{Phys. Rev. D}
  {\bfseries 18} (1978) 1829}.

\bibitem{Sikivie:1983ip}
P.~Sikivie, \emph{{Experimental Tests of the Invisible Axion}},
  \href{https://doi.org/10.1103/PhysRevLett.51.1415}{\emph{Phys. Rev. Lett.}
  {\bfseries 51} (1983) 1415}.

\bibitem{CAST:2017uph}
{\scshape CAST} collaboration, \emph{{New CAST Limit on the Axion-Photon
  Interaction}}, \href{https://doi.org/10.1038/nphys4109}{\emph{Nature Phys.}
  {\bfseries 13} (2017) 584}
  [\href{https://arxiv.org/abs/1705.02290}{{\ttfamily 1705.02290}}].

\bibitem{Biggio:2006im}
C.~Biggio, E.~Masso and J.~Redondo, \emph{{Mixing of photons with massive
  spin-two particles in a magnetic field}},
  \href{https://doi.org/10.1103/PhysRevD.79.015012}{\emph{Phys. Rev. D}
  {\bfseries 79} (2009) 015012}
  [\href{https://arxiv.org/abs/hep-ph/0604062}{{\ttfamily hep-ph/0604062}}].

\bibitem{Domcke:2025qlw}
V.~Domcke, C.~Garcia-Cely and S.M.~Lee, \emph{{Gravitational Wave Scattering on
  Magnetic Fields}},  \href{https://arxiv.org/abs/2507.16609}{{\ttfamily
  2507.16609}}.

\bibitem{Raffelt:1987im}
G.~Raffelt and L.~Stodolsky, \emph{{Mixing of the Photon with Low Mass
  Particles}}, \href{https://doi.org/10.1103/PhysRevD.37.1237}{\emph{Phys. Rev.
  D} {\bfseries 37} (1988) 1237}.

\bibitem{Gertsenshtein}
M.E.~Gertsenshtein, \emph{{ Wave Resonance of Light and Gravitational Waves }},
  {\emph{Sov. Phys. JETP} {\bfseries 14} (1962) 84}.

\bibitem{Boccaletti1970}
D.~Boccaletti, V.~De~Sabbata, P.~Fortini and C.~Gualdi, \emph{Conversion of
  photons into gravitons and vice versa in a static electromagnetic field},
  \href{https://doi.org/10.1007/BF02710177}{\emph{Il Nuovo Cimento B
  (1965-1970)} {\bfseries 70} (1970) 129}.

\bibitem{Ejlli:2019bqj}
A.~Ejlli, D.~Ejlli, A.M.~Cruise, G.~Pisano and H.~Grote, \emph{{Upper limits on
  the amplitude of ultra-high-frequency gravitational waves from graviton to
  photon conversion}},
  \href{https://doi.org/10.1140/epjc/s10052-019-7542-5}{\emph{Eur. Phys. J. C}
  {\bfseries 79} (2019) 1032}
  [\href{https://arxiv.org/abs/1908.00232}{{\ttfamily 1908.00232}}].

\bibitem{Domcke:2020yzq}
V.~Domcke and C.~Garcia-Cely, \emph{{Potential of radio telescopes as
  high-frequency gravitational wave detectors}},
  \href{https://doi.org/10.1103/PhysRevLett.126.021104}{\emph{Phys. Rev. Lett.}
  {\bfseries 126} (2021) 021104}
  [\href{https://arxiv.org/abs/2006.01161}{{\ttfamily 2006.01161}}].

\bibitem{vanDam:1970vg}
H.~van Dam and M.J.G.~Veltman, \emph{{Massive and massless Yang-Mills and
  gravitational fields}},
  \href{https://doi.org/10.1016/0550-3213(70)90416-5}{\emph{Nucl. Phys. B}
  {\bfseries 22} (1970) 397}.

\bibitem{Zakharov:1970cc}
V.I.~Zakharov, \emph{{Linearized gravitation theory and the graviton mass}},
  {\emph{JETP Lett.} {\bfseries 12} (1970) 312}.

\bibitem{Vainshtein:1972sx}
A.I.~Vainshtein, \emph{{To the problem of nonvanishing gravitation mass}},
  \href{https://doi.org/10.1016/0370-2693(72)90147-5}{\emph{Phys. Lett. B}
  {\bfseries 39} (1972) 393}.

\bibitem{Vinyoles:2016djt}
N.~Vinyoles, A.M.~Serenelli, F.L.~Villante, S.~Basu, J.~Bergstr\"om,
  M.C.~Gonzalez-Garcia et~al., \emph{{A new Generation of Standard Solar
  Models}}, \href{https://doi.org/10.3847/1538-4357/835/2/202}{\emph{Astrophys.
  J.} {\bfseries 835} (2017) 202}
  [\href{https://arxiv.org/abs/1611.09867}{{\ttfamily 1611.09867}}].

\bibitem{1954AuJPh...7..373S}
E.E.~{Salpeter}, \emph{{Electrons Screening and Thermonuclear Reactions}},
  \href{https://doi.org/10.1071/PH540373}{\emph{Australian Journal of Physics}
  {\bfseries 7} (1954) 373}.

\bibitem{Raffelt:1985nk}
G.G.~Raffelt, \emph{{ASTROPHYSICAL AXION BOUNDS DIMINISHED BY SCREENING
  EFFECTS}}, \href{https://doi.org/10.1103/PhysRevD.33.897}{\emph{Phys. Rev. D}
  {\bfseries 33} (1986) 897}.

\bibitem{Hoof:2021mld}
S.~Hoof, J.~Jaeckel and L.J.~Thormaehlen, \emph{{Quantifying uncertainties in
  the solar axion flux and their impact on determining axion model
  parameters}},
  \href{https://doi.org/10.1088/1475-7516/2021/09/006}{\emph{JCAP} {\bfseries
  09} (2021) 006} [\href{https://arxiv.org/abs/2101.08789}{{\ttfamily
  2101.08789}}].

\bibitem{Altherr:1992mf}
T.~Altherr and U.~Kraemmer, \emph{{Gauge field theory methods for
  ultradegenerate and ultrarelativistic plasmas}},
  \href{https://doi.org/10.1016/0927-6505(92)90014-Q}{\emph{Astropart. Phys.}
  {\bfseries 1} (1992) 133}.

\bibitem{Kapusta:1989tk}
J.I.~Kapusta, \emph{{Finite Temperature Field Theory, sec.~}}, Cambridge
  Monographs on Mathematical Physics, Cambridge University Press, Cambridge
  (1989).

\bibitem{Weinberg:1965nx}
S.~Weinberg, \emph{{Infrared photons and gravitons}},
  \href{https://doi.org/10.1103/PhysRev.140.B516}{\emph{Phys. Rev.} {\bfseries
  140} (1965) B516}.

\bibitem{Weinberg:1972kfs}
S.~Weinberg, \emph{{Gravitation and Cosmology}: {Principles and Applications of
  the General Theory of Relativity}}, no.~Sec.~10.4, John Wiley and Sons, New
  York (1972).

\bibitem{Gould:1985}
R.J.~{Gould}, \emph{{The graviton luminosity of the sun and other stars}},
  \href{https://doi.org/10.1086/162848}{\emph{apj} {\bfseries 288} (1985) 789}.

\bibitem{Steane:2023gme}
A.M.~Steane, \emph{{Gravitational bremsstrahlung in plasmas and clusters}},
  \href{https://arxiv.org/abs/2309.06972}{{\ttfamily 2309.06972}}.

\bibitem{Barker:1969jk}
B.M.~Barker, S.N.~Gupta and J.~Kaskas, \emph{{Graviton bremsstrahlung and
  infrared divergence}},
  \href{https://doi.org/10.1103/PhysRev.182.1391}{\emph{Phys. Rev.} {\bfseries
  182} (1969) 1391}.

\bibitem{Papini:1977fm}
G.~Papini and S.R.~Valluri, \emph{{Gravitons in Minkowski Space-Time.
  Interactions and Results of Astrophysical Interest}},
  \href{https://doi.org/10.1016/0370-1573(77)90006-0}{\emph{Phys. Rept.}
  {\bfseries 33} (1977) 51}.

\bibitem{Gould:1981an}
R.J.~Gould, \emph{{QUADRUPOLE BREMSSTRAHLUNG IN THE SCATTERING OF IDENTICAL
  CHARGED BOSONS AND FERMIONS}},
  \href{https://doi.org/10.1103/PhysRevA.23.2851}{\emph{Phys. Rev. A}
  {\bfseries 23} (1981) 2851}.

\bibitem{Voronov:1973kga}
N.A.~Voronov, \emph{{Gravitational Compton effect and photoproduction of
  gravitons by electrons}}, {\emph{Sov. Phys. JETP} {\bfseries 37} (1973) 953}.

\bibitem{Wadekar:2021qae}
D.~Wadekar and Z.~Wang, \emph{{Strong constraints on decay and annihilation of
  dark matter from heating of gas-rich dwarf galaxies}},
  \href{https://doi.org/10.1103/PhysRevD.106.075007}{\emph{Phys. Rev. D}
  {\bfseries 106} (2022) 075007}
  [\href{https://arxiv.org/abs/2111.08025}{{\ttfamily 2111.08025}}].

\bibitem{Todarello:2024qci}
E.~Todarello and M.~Regis, \emph{{Bounds on axions-like particles shining in
  the ultra-violet}},
  \href{https://doi.org/10.1088/1475-7516/2025/05/070}{\emph{JCAP} {\bfseries
  05} (2025) 070} [\href{https://arxiv.org/abs/2412.02543}{{\ttfamily
  2412.02543}}].

\bibitem{Chen:2014oda}
Y.J.~Chen, W.K.~Tham, D.E.~Krause, D.~Lopez, E.~Fischbach and R.S.~Decca,
  \emph{{Stronger Limits on Hypothetical Yukawa Interactions in the
  30{\textendash}8000 nm Range}},
  \href{https://doi.org/10.1103/PhysRevLett.116.221102}{\emph{Phys. Rev. Lett.}
  {\bfseries 116} (2016) 221102}
  [\href{https://arxiv.org/abs/1410.7267}{{\ttfamily 1410.7267}}].

\bibitem{Gondolo:2008dd}
P.~Gondolo and G.G.~Raffelt, \emph{{Solar neutrino limit on axions and keV-mass
  bosons}}, \href{https://doi.org/10.1103/PhysRevD.79.107301}{\emph{Phys. Rev.
  D} {\bfseries 79} (2009) 107301}
  [\href{https://arxiv.org/abs/0807.2926}{{\ttfamily 0807.2926}}].

\bibitem{ayala:2014pea}
A.~Ayala, I.~Dom{\'\i}nguez, M.~Giannotti, A.~Mirizzi and O.~Straniero,
  \emph{{Revisiting the bound on axion-photon coupling from Globular
  Clusters}}, \href{https://doi.org/10.1103/PhysRevLett.113.191302}{\emph{Phys.
  Rev. Lett.} {\bfseries 113} (2014) 191302}
  [\href{https://arxiv.org/abs/1406.6053}{{\ttfamily 1406.6053}}].

\bibitem{IAXO:2025ltd}
{\scshape IAXO} collaboration, \emph{{The International Axion Observatory
  (IAXO): case, status and plans. Input to the European Strategy for Particle
  Physics}},  \href{https://arxiv.org/abs/2504.00079}{{\ttfamily 2504.00079}}.

\bibitem{IAXO:2020wwp}
{\scshape IAXO} collaboration, \emph{{Conceptual design of BabyIAXO, the
  intermediate stage towards the International Axion Observatory}},
  \href{https://doi.org/10.1007/JHEP05(2021)137}{\emph{JHEP} {\bfseries 05}
  (2021) 137} [\href{https://arxiv.org/abs/2010.12076}{{\ttfamily
  2010.12076}}].

\bibitem{Galan:2015msa}
J.~Gal\'an et~al., \emph{{Exploring 0.1\textendash{}10 eV axions with a new
  helioscope concept}},
  \href{https://doi.org/10.1088/1475-7516/2015/12/012}{\emph{JCAP} {\bfseries
  12} (2015) 012} [\href{https://arxiv.org/abs/1508.03006}{{\ttfamily
  1508.03006}}].

\bibitem{Adelberger:2003zx}
E.G.~Adelberger, B.R.~Heckel and A.E.~Nelson, \emph{{Tests of the gravitational
  inverse square law}},
  \href{https://doi.org/10.1146/annurev.nucl.53.041002.110503}{\emph{Ann. Rev.
  Nucl. Part. Sci.} {\bfseries 53} (2003) 77}
  [\href{https://arxiv.org/abs/hep-ph/0307284}{{\ttfamily hep-ph/0307284}}].

\bibitem{Adelberger:2009zz}
E.G.~Adelberger, J.H.~Gundlach, B.R.~Heckel, S.~Hoedl and S.~Schlamminger,
  \emph{{Torsion balance experiments: A low-energy frontier of particle
  physics}}, \href{https://doi.org/10.1016/j.ppnp.2008.08.002}{\emph{Prog.
  Part. Nucl. Phys.} {\bfseries 62} (2009) 102}.

\bibitem{Decca:2005qz}
R.S.~Decca, D.~Lopez, H.B.~Chan, E.~Fischbach, D.E.~Krause and C.R.~Jamell,
  \emph{{Constraining new forces in the Casimir regime using the isoelectronic
  technique}}, \href{https://doi.org/10.1103/PhysRevLett.94.240401}{\emph{Phys.
  Rev. Lett.} {\bfseries 94} (2005) 240401}
  [\href{https://arxiv.org/abs/hep-ph/0502025}{{\ttfamily hep-ph/0502025}}].

\bibitem{Arkani-Hamed:1998sfv}
N.~Arkani-Hamed, S.~Dimopoulos and G.R.~Dvali, \emph{{Phenomenology,
  astrophysics and cosmology of theories with submillimeter dimensions and TeV
  scale quantum gravity}},
  \href{https://doi.org/10.1103/PhysRevD.59.086004}{\emph{Phys. Rev. D}
  {\bfseries 59} (1999) 086004}
  [\href{https://arxiv.org/abs/hep-ph/9807344}{{\ttfamily hep-ph/9807344}}].

\bibitem{Hannestad:2001jv}
S.~Hannestad and G.~Raffelt, \emph{{New supernova limit on large extra
  dimensions}},
  \href{https://doi.org/10.1103/PhysRevLett.87.051301}{\emph{Phys. Rev. Lett.}
  {\bfseries 87} (2001) 051301}
  [\href{https://arxiv.org/abs/hep-ph/0103201}{{\ttfamily hep-ph/0103201}}].

\bibitem{Hannestad:2001xi}
S.~Hannestad and G.G.~Raffelt, \emph{{Stringent neutron star limits on large
  extra dimensions}},
  \href{https://doi.org/10.1103/PhysRevLett.88.071301}{\emph{Phys. Rev. Lett.}
  {\bfseries 88} (2002) 071301}
  [\href{https://arxiv.org/abs/hep-ph/0110067}{{\ttfamily hep-ph/0110067}}].

\bibitem{Hannestad:2003yd}
S.~Hannestad and G.G.~Raffelt, \emph{{Supernova and neutron star limits on
  large extra dimensions reexamined}},
  \href{https://doi.org/10.1103/PhysRevD.69.029901}{\emph{Phys. Rev. D}
  {\bfseries 67} (2003) 125008}
  [\href{https://arxiv.org/abs/hep-ph/0304029}{{\ttfamily hep-ph/0304029}}].

\bibitem{Hardy:2025ajb}
E.~Hardy, A.~Sokolov and H.~Stubbs, \emph{{Stellar cooling limits on KK
  gravitons and dark dimensions}},
  \href{https://arxiv.org/abs/2510.18975}{{\ttfamily 2510.18975}}.

\end{thebibliography}\endgroup


\end{document}